\documentclass[aps,prl,showpacs,preprintnumbers,amssymb,superscriptaddress,twocolumn]{revtex4-1}
\usepackage[dvipdfmx]{graphicx}
\usepackage[dvipdfmx]{epsfig}
\usepackage{bm}
\usepackage{amsmath}
\usepackage{ascmac}
\usepackage{color}
\usepackage{setspace}
\usepackage{amssymb}
\usepackage{latexsym}
\usepackage{epstopdf}

\begin{document}

\title{Unconventional Universality Class of One-Dimensional Isolated Coarsening Dynamics in a Spinor Bose Gas} 

\author{Kazuya Fujimoto}
\affiliation{Department of Physics, University of Tokyo, 7-3-1 Hongo, Bunkyo-ku, Tokyo 113-0033, Japan}

\author{Ryusuke Hamazaki}
\affiliation{Department of Physics, University of Tokyo, 7-3-1 Hongo, Bunkyo-ku, Tokyo 113-0033, Japan}

\author{Masahito Ueda}
\affiliation{Department of Physics, University of Tokyo, 7-3-1 Hongo, Bunkyo-ku, Tokyo 113-0033, Japan}
\affiliation{RIKEN Center for Emergent Matter Science (CEMS), Wako, Saitama 351-0198, Japan}

\date{\today}

\begin{abstract}
By studying the coarsening dynamics of a one-dimensional spin-1 Bose-Hubbard model in a superfluid regime, we analytically find an unconventional universal dynamical scaling for the growth of the spin correlation length, which is characterized by the exponential integral unlike the conventional power-law or simple logarithmic behavior, and numerically confirmed with the truncated Wigner approximation.
\end{abstract}

\maketitle

$Introduction-$
Coarsening is a relaxation dynamics following a sudden change in system's parameters across a phase transition point. It has been studied in diverse classical systems of immense practical and fundamental importance such as magnetization processes, metal alloying, mixing of binary liquids, and nucleation in the gas-liquid transition \cite{HH,CD1}. The notable feature of coarsening is the dynamical scaling $C(r,t) = f(r/L_{\rm c}(t))$, which means that the correlation function $C(r,t)$ is characterized by a single length scale, namely the correlation length $L_{\rm c}(t)$ with a universal function $f(x)$.  The time dependence of $L_{\rm c}(t)$ classifies coarsening in various open dissipative systems described by, $e.g.,$ Ginzburg-Landau, Cahn-Hilliard equations into several universality classes that depend on basic information of systems such as spatial dimensions and symmetries. 

Recently, the relaxation dynamics including coarsening has attracted considerable attention in ultracold atomic gases which emerge as an ideal platform for studying nonequilibrium statistical mechanics in isolated quantum systems \cite{Pol11,Eisert15,Huse}. Indeed, over the last decade, many theoretical and experimental studies have uncovered a rich variety of relaxation phenomena in isolated quantum systems such as pre-thermalization \cite{prethermal1,prethermal2}, many-body localization \cite{Huse,MBL1,MBL2,MBL3}, transport dynamics \cite{DBrelaxation1,DBrelaxation2,DBrelaxation3,DBrelaxation4}, and the Kibble-Zurek mechanism (KZM) \cite{KZ1,KZ2,KZ3,KZ4}.

Then the following question naturally arises: ``Are there any unconventional universality classes unique to isolated coarsening dynamics?" Recently, coarsening dynamics in two-dimensional (2D) and three-dimensional (3D) multi-component Bose-Einstein condensates (BECs) have been investigated \cite{CDinBEC1,CDinBEC2,CDinBEC3,CDinBEC3_1,CDinBEC4,CDinBEC5,CDinBEC6,CDinBEC7,CDinBEC8}, which turn out to belong to the same conventional classes as in open dissipative systems such as the classical binary liquid and the planar spin model \cite{CDinBEC2,CDinBEC3,CDinBEC4,CDinBEC5,CDinBEC8}. As for the 2D coarsening dynamics with domains, this is due to the fact that the curvature and the inertia are the main driving forces promoting the coarsening both for 2D BECs and 2D classical binary liquids. In binary liquids, these forces overcome the effect of the dissipation in an inertial hydrodynamic regime, and the system effectively behaves as an isolated system, showing the characteristic power law $L_{\rm c}(t) \propto t^{2/3}$. The previous works for the 2D BECs \cite{CDinBEC2,CDinBEC3,CDinBEC4,CDinBEC5,CDinBEC8} confirmed this conventional universality class. Thus, it is still open whether the universality unique to isolated systems exists.

In this Letter, we theoretically investigate a one-dimensional (1D) spin-1 Bose-Hubbard (BH) model to demonstrate that the 1D isolated quantum system exhibits coarsening dynamics that belongs to an unconventional universality class. Unlike 2D and 3D systems, the curvature and the torsion of domain walls are absent in 1D systems, so that a 1D domain-wall interaction is generally weak. In open dissipative systems, such a genuine interaction between 1D topological objects is masked by the effect of dissipation \cite{comment1}; however, in 1D isolated systems it should become significant. More specifically, while a single 1D domain-wall pair is known to contract by itself in open dissipative systems \cite{KO,EO}, we find that in an isolated system such a pair undergoes a linear uniform motion without self-contraction. Based on this physical intuition, we obtain an analytical expression of $L_{\rm c}(t)$ characterized by an exponential integral, and numerically confirm it on the basis of the truncated Wigner approximation (TWA). This behavior is distinct from any power-law or simple logarithmic behavior found in open dissipative 1D systems \cite{1GL1,1GL2,1GL3,1CH1,1CH2,1CH3,1GD}, and attributed to the genuine interaction between the topological objects under energy conservation and to the absence of the curvature and torsion of a domain wall. Thus, the universality class found here is unique to 1D isolated systems. 
 
Some comments on previous related studies are in order here. The 1D domain-wall dynamics has been investigated numerically and experimentally in multi-component BECs, and short-time domain dynamics and the KZM have been discussed \cite{CDinBEC10,CDinBEC11,quench0,quench1,quench2,quench3,quench4}. However, universal coarsening behaviors such as a dynamical scaling have not been addressed. In contrast to the long-time coarsening dynamics, Nicklas $et~al.$ have focused on the short-time dynamics after the quench and experimentally investigated the universal dynamical scaling related to the KZM \cite{quench5}. Recently, Maraga $et~al.$ have studied coarsening in the $O(N)$ model and reported the breakdown of usual dynamical scaling \cite{On}; however, this result is not well understood from the perspective of universality classes.  

$Model-$
We consider a system of spin-1 bosons in a 1D optical lattice with a lattice constant $a$. 
Under the tight-binding approximation, this system is well described by the 1D spin-1 BH model \cite{SBH}.  Representing annihilation and creation operators of bosons with magnetic quantum number $m$ at the $j$th site as $b_{m,j}$ and $b^{\dagger}_{m,j}$ ($m=1,0,-1$), the Hamiltonian is given by
\begin{eqnarray}
\hat{H} = &-&J \sum_{m,j} \Bigl(\hat{b}_{m,j+1}^{\dagger} \hat{b}_{m,j} + {\rm h.c.} \Bigl) + q \sum _{m,j} m^2\hat{b}^{\dagger}_{m,j} \hat{b}_{m,j} \nonumber \\
&+& \frac{U_0}{2} \sum_{j} \hat{\rho}_j(\hat{\rho}_j-1) + \frac{U_2}{2} \sum_j \Bigl( \hat{\bm{S}}_j^2 - 2 \hat{\rho}_j \Bigl), \label{spinor_BH}
\end{eqnarray}
where $J$, $q$, $U_0$, and $U_2$ characterize the hopping amplitude, the quadratic Zeeman term, the density-dependent interaction, and the spin-dependent interaction, respectively. The operators for the total particle number and the spin vector at the $j$th site are defined by $\hat{\rho}_j = \sum_{m}  \hat{b}_{m,j}^{\dagger} \hat{b}_{m,j}$ and $\hat{S}_{\alpha,j}= \sum_{m,n}  \hat{b}_{m,j}^{\dagger} ({S_{\alpha}})_{mn} \hat{b}_{n,j}~(\alpha=x,y,z)$ with the spin-1 spin matrices $(S_{\alpha})_{mn}$. 

\begin{figure}[t]
\begin{center}
\includegraphics[keepaspectratio, width=8.5cm,clip]{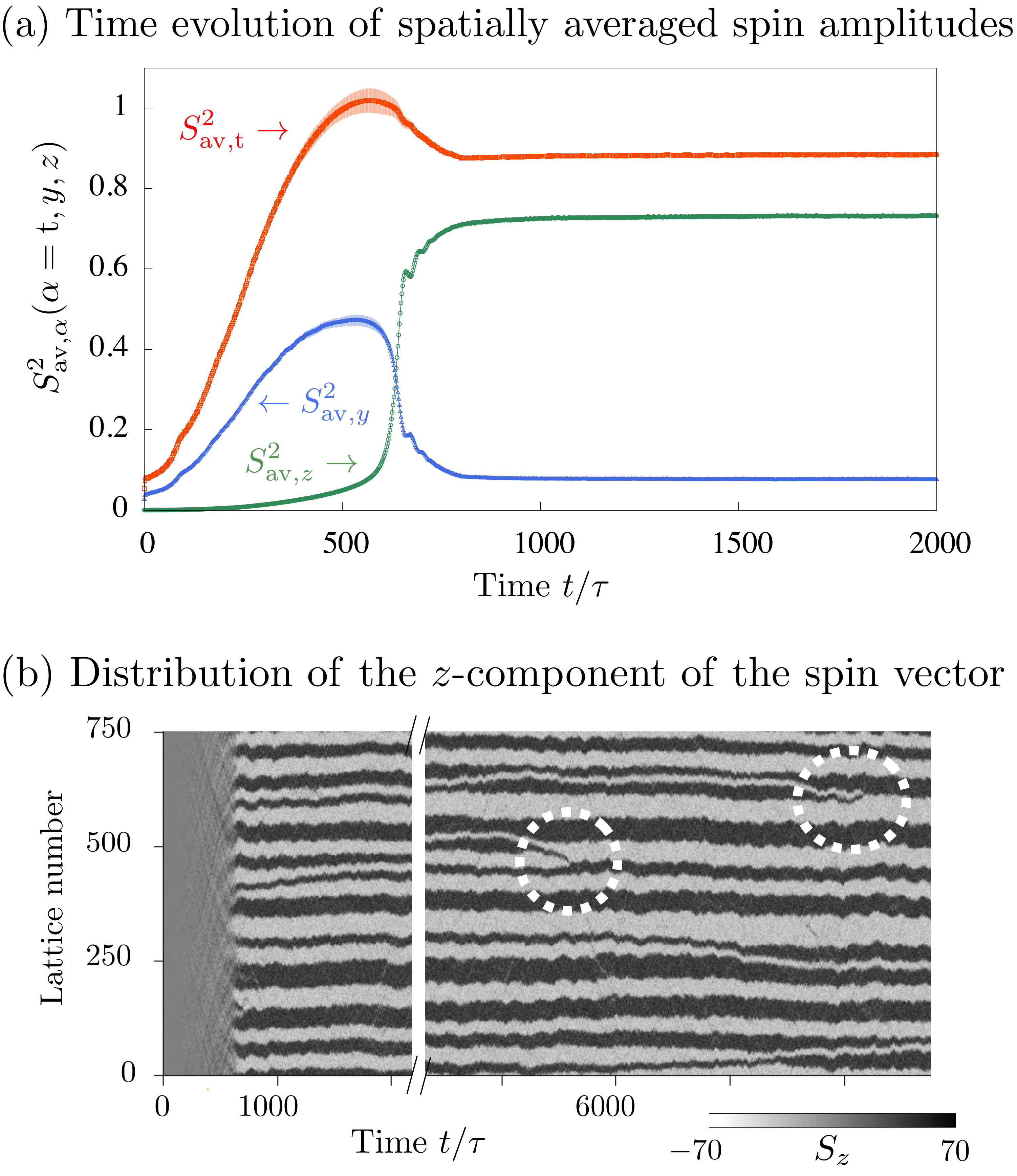}
\caption{(a) Time evolution of $S_{{\rm av},\alpha}^{2}~(\alpha={\rm t},y,z)$ for quench time $\tau_{\rm q}=800 \tau$. Color bands show $3\sigma/\sqrt{N_{\rm sam}}$ error bars in the TWA calculation, where $\sigma$ is the standard deviation and $N_{\rm sam}$ is the number of samples. The behavior of the $x$-component (not shown) is almost the same as that of the $y$-component. (b) Spatiotemporal distribution of $S_{z,j}$ corresponding to a single classical trajectory in the TWA calculation. Two dashed white circles indicate where domains merge.\label{spin_domain} }
\end{center}
\end{figure}

The ground state of this model is either a Mott-insulator phase or a superfluid phase depending on the parameters \cite{SBH}. In this work, we focus on a deep superfluid regime, where a dimensionless parameter $\kappa = \rho_{\rm f} J/U_0$ is much larger than unity. Here, $\rho_{\rm f} \equiv N/3M$ is the filling factor with the total particle number $N$ and the number of lattice points $M$.

$Numerical~result-$
We apply the TWA method \cite{Bla,Pol} to study the relaxation dynamics dominated by many spin domains. This method can incorporate effects of quantum fluctuations through sampling of initial states. The system is assumed to have a ferromagnetic interaction ($U_2<0$), and the parameters in Eq.~(\ref{spinor_BH}) are set to be $U_{0}/J=1/40$, $U_{2}/U_0=-1/10$, $N=40000$, and $M=1024$. Then, $\kappa$ is about $520$ and the system is in a deep superfluid regime. The detailed numerical implementation is described in \cite{Supl}, where we demonstrate that, in a deep superfluid regime, TWA results find good agreement with results obtained by directly solving the Schr$\rm \ddot{o}$dinger equation with the Crank-Nicolson method \cite{CN}. 

\begin{figure}[t]
\begin{center}
\includegraphics[keepaspectratio, width=8.7cm,clip]{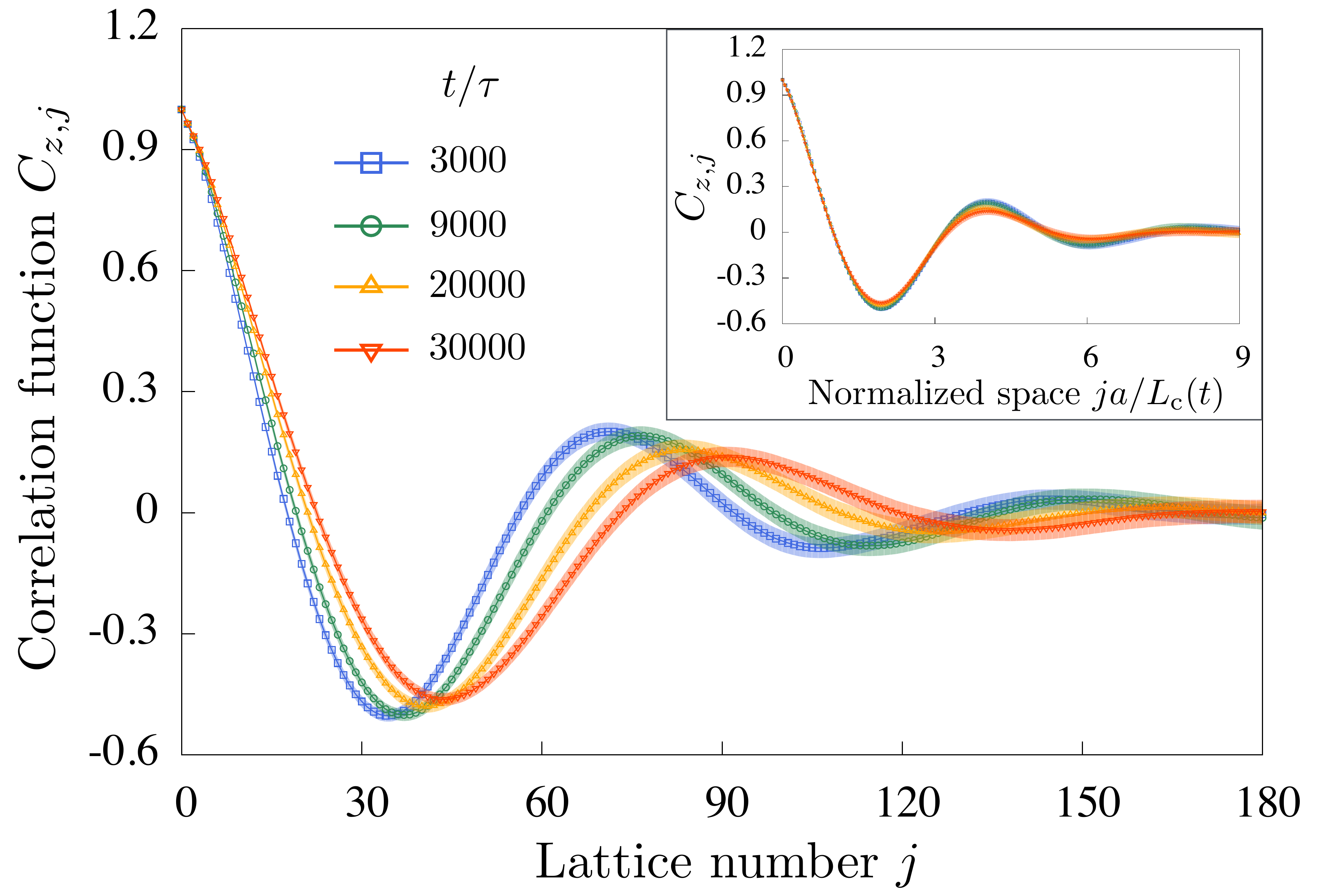}
\caption{Time evolution of the correlation function $C_{z,j}$ with $t/\tau=3000,9000,20000$, and $30000$. Color bands show $3\sigma/\sqrt{N_{\rm sam}}$ error bars. The first zero-crossing point of each curve defines the correlation length $L_{\rm c}(t)$. The inset shows $C_{z,j}$ with the $x$-axis normalized by $L_{c}(t)$. All curves converge to a single one.  
\label{spin_correlation} }
\end{center}
\end{figure}

To excite many spin domains, we quench the coefficient $q(t)$ for the Zeeman term as
\begin{eqnarray}
q(t)=\left\{ \begin{array}{ll}
-2.4 n U_{2} \bigl[ 1-(1+\frac{1}{2.4})\frac{t}{\tau_{q}} \bigl] & (t<\tau_{\rm q}); \\
n U_{2} & (\tau_{\rm q} \leq t), \\
\end{array} \right.
\end{eqnarray}
where $\tau_{\rm q}$ is the quench time and $n=N/M$. We choose $q(0)$ such that the initial state is a polar phase. This quench protocol crosses two phase-transition points  from the polar phase to the broken-axisymmetry phase and then to the ferromagnetic phase \cite{SBH}.

Figure~\ref{spin_domain} (a) shows the time evolution of the spin amplitude defined by $S_{{\rm av},\alpha}^{2}(t)=  \langle \sum _{j}\hat{S}_{\alpha,j}^2/n^2M \rangle (t) ~(\alpha=x,y,z)$ and $S_{{\rm av,t}} ^{2}(t)=  \sum_{\alpha} S_{{\rm av},\alpha}^2(t) $, where the bracket means a quantum average $\langle \cdots \rangle (t) = \langle \psi(t) | \cdots | \psi(t) \rangle$ with the state vector $| \psi(t) \rangle$ at time $t$. In this result, the quench time is set to be $\tau_{\rm q}=800\tau$ with $\tau=4\hbar/J$. At an early stage of the quench protocol, the $x$- and $y$-components of the spin vector rapidly grow because the system is brought to the broken-axisymmetry phase where the dynamical instabilities of the $m=\pm 1$ components lead to the increase of the particle number of those components. At a later stage, the instability of the $m=0$ component becomes strong as the system enters the ferromagnetic phase. Then, the particle number of the $m=0$ component rapidly decreases and eventually the $z$-component dominates the other components.

After the quench, many domain walls are formed as shown in Fig.~\ref{spin_domain} (b), which is a spatiotemporal distribution of $S_{z,j}$ obtained by a single classical trajectory of the TWA calculation. The encircled regions shows where spin domains merge. This merging process enlarges domain structures. 

To investigate this coarsening behavior quantitatively, we calculate a spatial correlation function $C_{z,j}(t)$ for $\hat{S}_{z,i}$ defined by 
\begin{eqnarray}
C_{z,j}(t) = \frac{\sum_{k=1}^{M} \langle \hat{S}_{z,j+k}\hat{S}_{z,k} \rangle(t)}{\sum_{k=1}^{M} \langle \hat{S}_{z,k}\hat{S}_{z,k} \rangle(t)}.
\end{eqnarray}
Figure \ref{spin_correlation} shows the time evolution of $C_{z,j}$, and the inset shows the same curves with the abscissa normalized by the correlation length $L_{\rm c}(t)$ defined by the first zero crossing point of the correlation function. We find that all curves are rescaled into a single universal curve, showing a dynamical scaling characteristic of coarsening dynamics. A small deviation from the single curve is expected to be caused by density and spin waves excited by merging of the domains, which cannot dissipate in the isolated system and weaken long-range correlations. Actually, in dissipative 1D systems, clear dynamical scaling without a small deviation has been confirmed \cite{DS1,DS2}.

\begin{figure}[t]
\begin{center}
\includegraphics[keepaspectratio, width=8.5cm,clip]{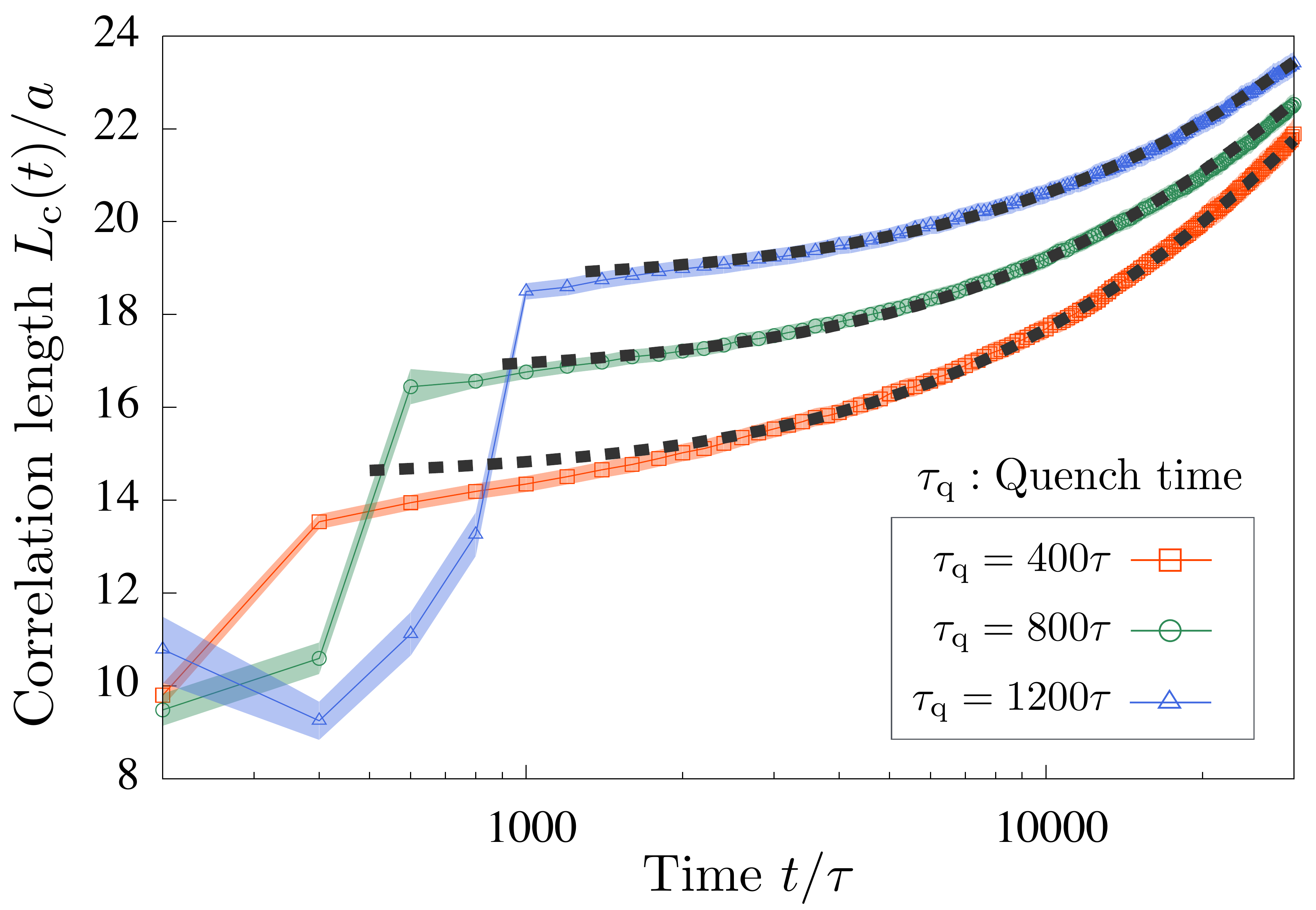}
\caption{Time evolution of the correlation length $L_{\rm c}(t)$ for quench times $\tau_{\rm q}/\tau=400,800$, and $1200$. Color bands show error bars. Black dashed curves show analytic results of Eq.~(\ref{domain_law3}). \label{domain_size} }
\end{center}
\end{figure}

To understand the universality class, we examine the time evolution of $L_{\rm c}(t)$. Figure \ref{domain_size} shows $L_{\rm c}(t)$, which exhibits behavior quite different from the conventional logarithmic and power laws \cite{1GL1,1GL2,1GL3,1CH1,1CH2,1CH3,1GD}. 

$Analytic~result-$
We show that the growth law of $L_{\rm c}(t)$ in Fig.~\ref{domain_size} is characterized by the exponential integral. As shown in Fig.~\ref{spin_domain}(b), the size of a domain grows only through merging of two domain-wall pairs. This suggests that a domain-wall pair plays a key role here.

To analyze the domain-wall pair dynamics, we note that this system can be transformed to a continuum model similar to the spinor Gross-Pitaevskii equation because the width of the domain wall $2\lambda=2a\sqrt{J/|q_{\rm F}|}=6.4a$ at $q_{\rm F}=q(\tau_{\rm q})$ is larger than the lattice constant $a$. Then, we can derive a spin-hydrodynamic equation  \cite{spin_hydro1,spin_hydro2,spin_hydro3,spin_hydro4}, $i.e.$ the Landau-Lifshitz (LL) equation, given by 
\begin{eqnarray}
\frac{\partial}{\partial t} {\bm S}(x,t) = -\bm{S}(x,t) \times \bm{B}(x,t),  \label{1LL1}
\end{eqnarray}
where $\bm{B}(x,t) = -J'\nabla^2 \bm{S}(x,t) - q' S_z\hat{\bm e}_z$, $J'=Ja^2/\hbar$ and $q'=-q_{\rm F}/\hbar$. The derivation of Eq.~(\ref{1LL1}) is described in \cite{Supl}.

Applying the singular perturbation method \cite{SP1,KO,EO} to Eq.~(\ref{1LL1}), we find that a domain-wall pair undergoes a linear uniform motion with velocity $V(l)$ given by 
\begin{eqnarray}
V(l) =  4\sqrt{J'q'} e^{-l/\lambda} {\rm sin}(\phi_1-\phi_2).  \label{velocity}
\end{eqnarray}
Here, $l$ is the distance between the two domain walls, and the phase $\phi_j$ is the azimuthal angle of $\bm{S}_{j}$ at the center of the domain wall labeled by $j=1,2$. The derivation is described in \cite{Supl}, where Eq.~(\ref{velocity}) is compared with numerical results of Eq.~(\ref{1LL1}).

Next, we investigate the correlation length $L_{\rm c}(t)$ by assuming that many domain-wall pairs randomly move and that merging of domains occurs through collisions between the domain-wall pairs. Let us examine a situation where there are $N_{\rm d}(t)$ domain walls, and the average distance between the walls at time $t$ is denoted as $l_{\rm d}(t)$. Firstly, we note that the average collision time $\tau_{\rm c}(t)$ is given by $l_{\rm d}(t)/V_{\rm av}(t)$. Here the average velocity is represented as $V_{\rm av}(t) = V_0 {\rm exp}(-l_{\rm d}(t)/\lambda)$ because of Eq.~(\ref{velocity}). However, we cannot determine the proportionality constant $V_0$ since the distribution of $\phi_1-\phi_2$ is complicated. Then, assuming only two properties of a Poisson process \cite{PP}, we derive the time derivative of $N_{\rm d}(t)$:
\begin{eqnarray}
\frac{d}{dt}N_{\rm d}(t) = -\frac{\alpha}{\tau_{\rm c}(t)} = -\frac{\alpha V_{\rm av}(l_{\rm d}(t))}{l_{\rm d}(t)}, \label{domain_law1}
\end{eqnarray}
where $\alpha$ is a positive constant. 

Secondly, we use the fact that $N_{\rm d}(t)$ is inversely proportional to $l_{\rm d}(t)$, which leads to $N_{\rm d}(t)=\beta /l_{\rm d}(t)$ with a positive constant $\beta$. We substitute it into Eq.~(\ref{domain_law1}), obtaining 
\begin{eqnarray}
\frac{d}{dt}l_{\rm d}(t) =  \gamma l_{\rm d}(t) {\rm exp}(-l_{\rm d}(t)/\lambda), \label{domain_law2}
\end{eqnarray}
where $\gamma = \alpha V_0 /\beta$. 

The solution to this equation is expressed by the exponential integral ${\rm Ei}[x] = \int_{-\infty}^{x} {\rm exp}(t)/t~dt$ \cite{ei}. Using this function and the fact that $l_{\rm d}(t)$ is proportional to the correlation length $L_{\rm c}(t) = \eta l_{\rm d}(t)$ with the proportionality constant $\eta$, we arrive at  
\begin{eqnarray}
L_{\rm c}(t) = \eta \lambda {\rm Ei}^{-1}[ \gamma (t-t_0)+ {\rm Ei}[ L_{\rm c}(t_0)/\eta\lambda]]. \label{domain_law3}
\end{eqnarray}
Here ${\rm Ei}^{-1}[b]=a$ is the inverse function of ${\rm Ei}[a] = b$. 
We note that this law asymptotically approaches a logarithmic law after a sufficiently long time. 

In Fig.~\ref{domain_size}, we plot this function as dashed curves, which are in excellent agreement with the numerical results. The deviations of the data ($\tau_{\rm q}/\tau=400$) in the early time are due to partial breakdown of Eq.~(\ref{velocity}) because it becomes a good approximation only when the distance $l$ is large. Note that Eq.~(\ref{domain_law3}) has two constants $\gamma$ and $\eta$. In Fig.~\ref{domain_size}, we use $(\gamma \tau,\eta) = (0.000380,1.72), (0.000385,1.70), (0.000380,1.69)$ for $\tau_{\rm q}/\tau=400,800,1200$, respectively. Thus, all numerical data can be fitted by almost  the same $\gamma$ and $\eta$. 

\begin{figure}[t]
\begin{center}
\includegraphics[keepaspectratio, width=8.5cm,clip]{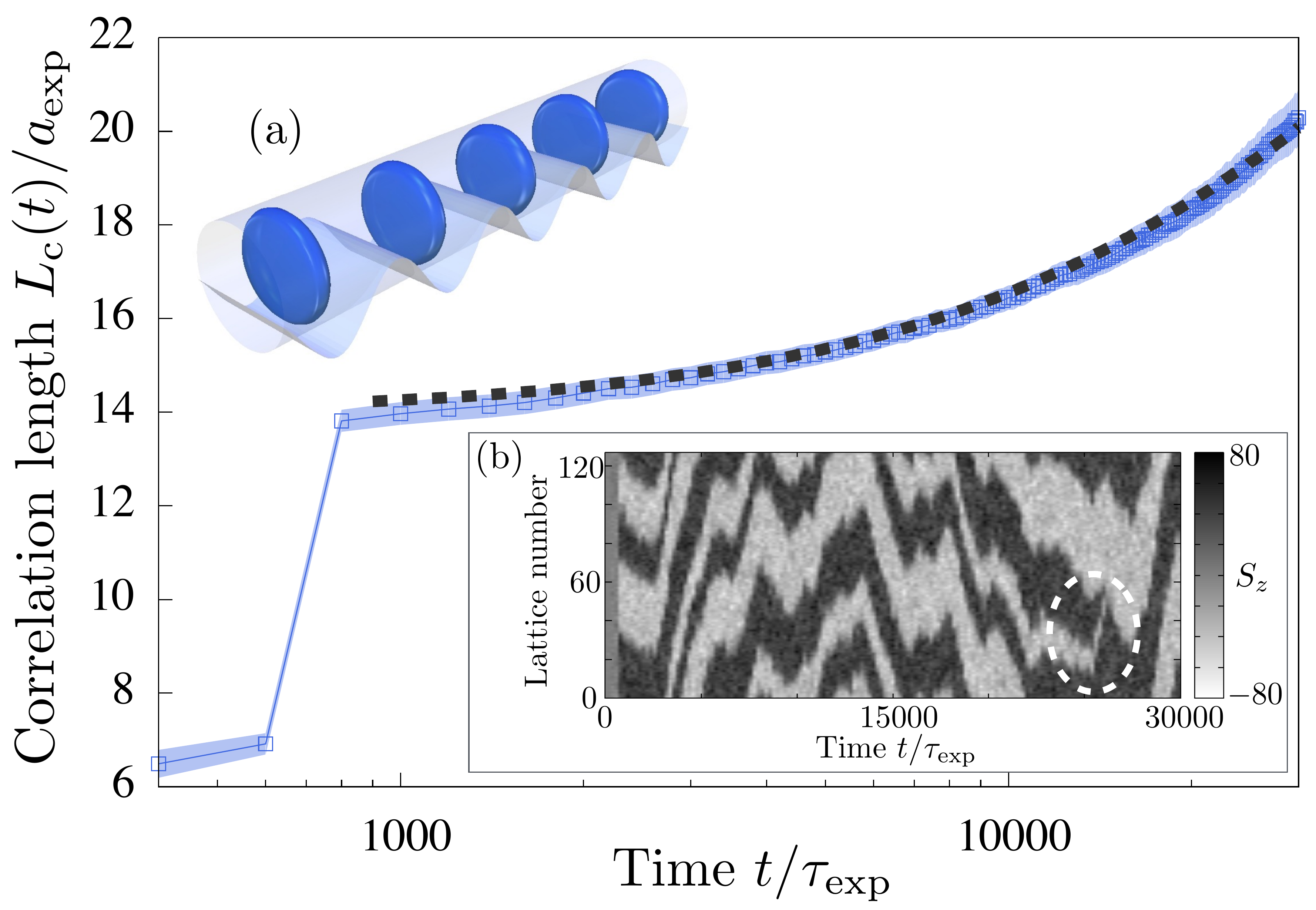}
\caption{Time evolution of the correlation length $L_{\rm c}(t)$ in the experimental setup. A black dashed curve is the analytic result of Eq.~(\ref{domain_law3}) with $(\gamma \tau_{\rm exp},\eta) = (0.000280,1.70)$, where a color band shows an error bar. Inset. (a) Possible experimental setup. (b) Spatiotemporal distribution of $S_{z,j}$ for a single classical trajectory in the TWA calculation.
\label{domain_size_exp} }
\end{center}
\end{figure}

Finally, we comment on the relation between our result of Eq.~(\ref{domain_law3}) and the previous works concerning the 1D coarsening \cite{1GL1,1GL2,1GL3,1CH1,1CH2,1CH3,1GD}. In these works, the energy is dissipated, so that two domain walls forming a wall pair contract by itself. Such a self-contraction was confirmed in Refs.~\cite{KO,EO}, but it is different from the coarsening process (merging of two domain-wall pairs) of our study. Thus, these systems do not obey Eq.~(\ref{domain_law3}), although both previous studies and ours show the same logarithmic behavior in the long-time limit. As an exception, there is a convective Cahn-Hilliard equation, where a domain-wall pair undergoes a linear uniform motion \cite{1CH3,CCH}. Thus we expect that this system obeys Eq.~(\ref{domain_law3}), though it was not derived in previous literature.

$Discussion-$
We first discuss why Eq.~(\ref{domain_law3}) is universal. As can be seen from the derivation, this law originates from the mechanism where a domain-wall pair moves at an average velocity proportional to ${\rm exp}(-l/\lambda)$ without the self-contraction. The exponential dependence of the velocity on $l$ and the absence of self-contraction are due to the interaction between the 1D domain walls and isolation from the environment \cite{reason}. Thus, Eq.~(\ref{domain_law3}) reflects the nature of a 1D isolated system. A typical example that satisfies these conditions is the 1D LL equation, which is a universal effective equation in 1D spin systems. Thus, we expect that the growth law of Eq.~(\ref{domain_law3}) is universal in a 1D isolated spin system if the domain is stable and the domain merging occurs. 

Next, we discuss possible experimental situations. A difficulty of observing Eq.~(\ref{domain_law3}) is a limited lifetime of trapped atoms. In a 1D system, the interaction between domain walls is weak due to the exponential spin configuration, so that the relaxation time is very long. However, Eq.~(\ref{domain_law3}) may be observed if we prepare a 1D system with $^{7}{\rm Li}~(F=1)$. 

We consider quasi-1D systems of $^{7}{\rm Li}$ atoms in a 1D optical lattice, where atoms are tightly trapped in a radial direction as shown in the inset (a) of Fig.~\ref{domain_size_exp}. The parameters used are $a_{\rm exp}=0.387~{\rm \mu m}$ \cite{lattice_constant}, $M_{\rm exp}=128$, $N_{\rm exp}=5000$, the radial trapping frequency $\omega_{\rm r}=2\pi \times 4500 {\rm /s}$, and the depth of the lattice $V_{\rm d}=5 E_{\rm r}$ with $E_{\rm r} = \hbar^2/8Ma_{\rm exp}^2$ being the recoil energy. Then, this system behaves as a quasi-1D system and a 1D calculation can be justified since the condition $\hbar \omega_{\rm r} \sim 2.8 \mu$ with the chemical potential $\mu$ is satisfied and excitations in the radial direction are suppressed. 

Under the above setup, we have performed a TWA calculation for the 1D system and confirmed Eq.~(\ref{domain_law3}) as shown in Fig.~\ref{domain_size_exp}. The characteristic time $\tau_{\rm exp}$ is about $0.156 ~{\rm ms}$, and the calculation terminates at about $4.7~{\rm s}$, which is accessible in current experiments. 
To measure the correlation function, we need the spatial resolution of about $1~{\rm \mu m}$, which is available in an in-situ imaging method \cite{quench5}.
When experiments continue until $9~{\rm s}$, we can obtain 4 data points for $L_{\rm c}(t)$ \cite{SpRe}. In this time evolution, $L_{\rm c}(t)$ is completely different from any power and logarithmic laws. Thus, we can distinguish Eq.~(\ref{domain_law3}) from the conventional laws.

Finally we discuss a finite-size effect and three-body loss. As for the former, we note that the number of domain walls is not large as shown in the inset (b) of Fig.~\ref{domain_size_exp}. Thus, in the long-time dynamics, the coarsening should be suppressed. However, we confirm Eq.~(\ref{domain_law3}) until about $4.7~{\rm s}$. After this time, the finite-size effect may be significant. As for the latter, the central density at each site is about $2.23 \times 10^{14}~{/\rm cm^3}$. Thus, if a three-body loss rate of $^{7}{\rm Li}$ is $6 \times 10^{-31} {\rm cm^6/s}$ \cite{TBL}, the particle loss until $9~{\rm s}$ is about $19 \%$, which allows experimental test of our predictions.

$Conclusion-$
The relaxation dynamics described by the 1D spin-1 BH model has been analytically and numerically studied. Our numerical calculation based on the TWA method has revealed that the system in a deep superfluid regime exhibits coarsening with the dynamical scaling that belongs to the universality class different from conventional classes. We have analytically obtained the universal domain-growth law of Eq.~(\ref{domain_law3}), which is in remarkable agreement with the numerical data. 

$Acknowledgements-$
We are grateful to  I. Danshita, S. Furukawa, I. Ichinose, S. Inouye, K. Kasamatsu, Y. Kuno, T. Matsui, and H. Takeuchi for useful discussion and comments on this work. This work was supported by KAKENHI Grant No. JP26287088 from the Japan Society for the Promotion of Science, and a Grant-in-Aid for Scientific Research on Innovative Areas “Topological Materials Science” (KAKENHI Grant No. JP15H05855), and the Photon Frontier Network Program from MEXT of Japan, ImPACT Program of Council for Science, Technology and Innovation (Cabinet Office, Government of Japan). K. F. was supported by JSPS fellowship (JSPS KAKENHI Grant No. JP16J01683). R. H. was supported by the Japan Society for the Promotion of Science through Program for Leading Graduate Schools (ALPS) and JSPS fellowship (JSPS KAKENHI Grant No. JP17J03189).

\widetext
\clearpage

\setcounter{equation}{0}
\setcounter{figure}{0}
\renewcommand{\theequation}{S-\arabic{equation}}
\renewcommand{\thefigure}{S-\arabic{figure}}

\section*{Supplemental Materials}
In this Supplemental Material, we discuss here the numerical implementation of the truncated Wigner approximation (TWA) calculation, and the dynamics of a spin domain-wall pair described by a one-dimensional (1D) Landau-Lifshitz (LL) equation. 

\section{Detailed numerical method for the TWA calculation}
We explain how the TWA method is applied to the spin-1 Bose-Hubbard (BH) model, and discuss the validity of the TWA method by comparing TWA and full quantum dynamics (FQD) results. Details of the TWA formulation is reviewed in Refs. \cite{Bla_S,Pol_S}. 

\subsection{Numerical implementation of the TWA method in the spin-1 BH model}
In the TWA calculation, we solve a classical equation corresponding to the spin-1 BH model, which is given by
\begin{eqnarray}
i\hbar \frac{\partial }{\partial t} b_{m,j} = \frac{\partial }{\partial b_{m,j}^*} \mathcal{H}_{\rm W}, \label{CE}
\end{eqnarray}
where the variables $b_{m,j}$ and $b_{m,j}^*$ are the c-numbers corresponding to the annihilation and creation operators, respectively. The Hamiltonian $\mathcal{H}_{\rm W}$ for the c-numbers is the Weyl representation of the Hamiltonian of the spin-1 BH model: 
\begin{eqnarray}
\mathcal{H}_{\rm W} =  & - & J \sum_{m,j} \Bigl(b_{m,j+1}^{*} b_{m,j} + b_{m,j}^{*} b_{m,j+1} \Bigl) + q \sum _{m,j} \Bigl( m^2  |b_{m,j}|^2 - \frac{1}{2}\Bigl) \nonumber \\  
 & + & \frac{U_0}{2} \sum_{j} ( \rho_j^2 - 4\rho_j + 3) + \frac{U_2}{2} \sum_j \Bigl( \bm{S}_j^2 - 2 \rho_j + \frac{3}{2} \Bigl).  \label{WHam}
\end{eqnarray}
The total particle number and spin vector at the $j$-site are defined by $\rho_j = \sum_{m}  |b_{m,j}|^2$ and ${\bm S}_j= \sum_{m,n}  b_{m,j}^{*} (\hat{\bm{S}})_{mn} b_{n,j}$, respectively, where $(\hat{\bm{S}})_{mn}$ are spin-1 matrices defined by 
\begin{alignat}{3}
\hat{S}_x=\frac{1}{\sqrt{2}}
\begin{pmatrix} 
0 & 1 & 0\\
1 & 0 & 1\\
0 & 1 & 0
\end{pmatrix},
\qquad
\hat{S}_y=\frac{i}{\sqrt{2}}
\begin{pmatrix} 
0 & -1 & 0\\
1 & 0 & -1\\
0 & 1 & 0
\end{pmatrix},
\qquad
\hat{S}_z=
\begin{pmatrix} 
1 & 0 & 0\\
0 & 0 & 0\\
0 & 0 & -1
\end{pmatrix}.
\end{alignat}
In the TWA method, the initial state is sampled from a Wigner function corresponding to an initial quantum state. This function contains information about quantum fluctuations in the initial state. In this sense, the TWA is an approximation beyond the mean field theory. 

In our study, the initial state is sampled from a Wigner function corresponding to the Bogoliubov vacuum $|0 \rangle _{\rm B}$ for the polar phase. This vacuum is defined by $\hat{a}_{m,k}|0 \rangle _{\rm B} = 0~(m=1,0,-1; k\in {\rm 1st BZ})$, where the operator $\hat{a}_{m,k}$ is an annihilation operator for the Bogoliubov quasiparticle, which is defined through Eqs.~(\ref{Bogo1})-(\ref{Bogo3}) below. The notation $k\in {\rm 1st BZ}$ means that wave number belongs to the 1st Brillouin zone. In the Bogoliubov theory, we assume that an operator corresponding to a condensate mode is replaced by a c-number and that the condensate fraction is set to be $N_0$. Then, the annihilation operator $\hat{b}_{m,j}$ can be expanded as 
\begin{eqnarray}
\hat{b}_{1,j} =  \frac{1}{\sqrt{M}}\sum_{ \substack{k\neq 0 \\ k \in {\rm 1st BZ} } } \biggl( u_{1,k}\hat{a}_{1,k} + v_{-1,k}\hat{a}_{-1,-k}^{\dagger} \biggl) {\rm exp} \biggl(i \frac{2\pi k j}{M} \biggl), \label{Bogo1}
\end{eqnarray}
\begin{eqnarray}
\hat{b}_{0,j} = \sqrt{N_0} + \frac{1}{\sqrt{M}}\sum_{ \substack{k\neq 0 \\ k \in {\rm 1st BZ} } } \biggl( u_{0,k}\hat{a}_{0,k} + v_{0,k}\hat{a}_{0,-k}^{\dagger} \biggl) {\rm exp} \biggl(i \frac{2\pi k j}{M} \biggl), \label{Bogo2} 
\end{eqnarray}
\begin{eqnarray}
\hat{b}_{-1,j} =  \frac{1}{\sqrt{M}}\sum_{ \substack{k\neq 0 \\ k \in {\rm 1st BZ} } } \biggl( u_{-1,k}\hat{a}_{-1,k} + v_{1,k}\hat{a}_{1,-k}^{\dagger} \biggl) {\rm exp} \biggl(i \frac{2\pi k j}{M} \biggl), \label{Bogo3}
\end{eqnarray}
where the coefficients $u_{m,k}$ and $v_{m,k}$ are given by
\begin{eqnarray}
u_{1,k}=u_{-1,k}=\sqrt{ \frac{\mathcal{A}_k+\mathcal{E}_{1,k}}{2\mathcal{E}_{1,,k}} },  \label{Bogo4}
\end{eqnarray}
\begin{eqnarray}
v_{1,k}=v_{-1,k}=\sqrt{ \frac{\mathcal{A}_k-\mathcal{E}_{1,k}}{2\mathcal{E}_{1,k}} },  \label{Bogo5}
\end{eqnarray}
\begin{eqnarray}
u_{0,k}= \sqrt{ \frac{\mathcal{B}_k+\mathcal{E}_{0,k}}{2\mathcal{E}_{0,k}} },  \label{Bogo6}
\end{eqnarray}
\begin{eqnarray}
v_{0,k}= -\sqrt{ \frac{\mathcal{B}_k-\mathcal{E}_{0,k}}{2\mathcal{E}_{0,k}} },  \label{Bogo7}
\end{eqnarray}
\begin{eqnarray}
\mathcal{A}_k = \epsilon_k + q + \frac{U_2N}{M}, 
\end{eqnarray}
\begin{eqnarray}
\mathcal{B}_k = \epsilon_k + \frac{U_0N}{M}, 
\end{eqnarray}
\begin{eqnarray}
\mathcal{E}_{1,k} = \sqrt{ \Big( \epsilon_k+q \Big) \Big( \epsilon_k +q+ \frac{2U_2N}{M} \Big) },
\end{eqnarray}
\begin{eqnarray}
\mathcal{E}_{0,k} = \sqrt{ \epsilon_k \Big( \epsilon_k + \frac{2U_0N}{M} \Big) },
\end{eqnarray}
with the number of lattice points $M$ and the dispersion relation $\epsilon_k = 2J-2J{\rm cos}({2\pi k }/{M})$ for free particles. 
Then, we can derive the Wigner function for the vacuum $|0 \rangle _{\rm B}$:
\begin{eqnarray}
W_{\rm B}( \{ a_{m,k}^* \}, \{ a_{m,k} \} ) = \prod_{m=-1}^{1} \prod_{ \substack{k\neq 0 \\ k \in {\rm 1st BZ} }  } \frac{2}{\pi}{\rm exp} \biggl( -2|a_{m,k}|^2 \biggl), \label{Wigner}
\end{eqnarray}
where $a_{m,k}^*$ and $a_{m,k}$ are the c-numbers corresponding to the operators $\hat{a}_{m,k}^{\dagger}$ and $\hat{a}_{m,k}$.
This function satisfies the normalization condition 
\begin{eqnarray}
\int W_{\rm B}( \{ a_{m,k}^* \}, \{ a_{m,k} \} )  \prod_{m=-1}^{1} \prod_{ \substack{k\neq 0 \\ k \in {\rm 1st BZ} }  } da_{m,k}^*da_{m,k}  = 1, 
\end{eqnarray}
where a measure $da_{m,k}^*da_{m,k} = d({\rm Re}[a_{m,k}]) d({\rm Im}[a_{m,k}])$. 

Note that in the above sampling the number of condensate particles is fixed to be $N_0$ as shown in Eq.~(\ref{Bogo2}). Alternatively, we may use a coherent state with the average particle number $N_0$. In this case, $a_{m,0}$ and $a^*_{m,0}$ are sampled from 
\begin{eqnarray}
W_{\rm C}( \{ a_{m,0}^* \}, \{ a_{m,0} \} ) = \prod_{m=-1}^{1}  \frac{2}{\pi}{\rm exp} \biggl( -2|a_{m,0}-\sqrt{N_0} \delta_{m,0}|^2 \biggl).
\end{eqnarray}
Our numerical calculation shows that the TWA calculations with these two ways exhibit quantitatively the same results. This is due to a large number of condensate particles with $N_0=40000$. Thus, in the main text, we show the data sampled from Eq.~(\ref{Wigner}) with the fixed $N_0$.

In this setup, we can perform the TWA calculation by computing $N_{\rm sam}$ classical solutions of Eq.~(\ref{CE}) with different initial states sampled from Eq.~(\ref{Wigner}). Firstly, to generate an initial state, we sample $a_{m,k}$ from Eq.~(\ref{Wigner}), and transform $a_{m,k}$ into $b_{m,k}$ by using a classical counterpart of Eqs.~(\ref{Bogo1})-(\ref{Bogo3}). Secondly, the classical equation of (\ref{CE}) is solved with this initial state. Repeating this procedure with different initial states, we obtain $N_{\rm sam}$ classical solutions $b_{m,j}^{(\alpha)} (t) ~ (\alpha=1,2,\cdots,N_{\rm sam})$. Then, a quantum average for an operator $\hat{A}$ is computed by 
\begin{eqnarray}
\langle \hat{A} \rangle(t) = \frac{1}{N_{\rm sam}} \sum_{\alpha=1}^{N_{\rm sam}} A_{\rm W}( \{ b_{m,j}^{(\alpha)^*} (t) \}, \{ b_{m,j}^{(\alpha)} (t) \}).
\end{eqnarray}
Here, $A_{\rm W}( \{ b_{m,j}^{(\alpha)^*} \}, \{ b_{m,j}^{(\alpha)} \})$ is the Weyl representation of $\hat{A}$, which is derived by the same way as in Eq.~(\ref{WHam}). 

We comment on the validity of the TWA method. In the TWA formulation \cite{Pol_S}, we assume that quantum fluctuations from the classical trajectory obeying a mean field equation are small. Therefore, this method becomes a good approximation if the parameters of the BH model are set to be in the deep superfluid regime ($\rho_{\rm f}J/U_0 \gg 1$ and $\rho_{\rm f} \gg 1$) and the system size is not so large, where $\rho_{\rm f}$ is the filling factor. The latter condition reflects the Mermin-Wagner-Hohenberg theorem \cite{BEC1_S}, which states that the mean field prediction is wrong in the thermodynamic limit. For this reason, the system size is determined by the condition such that the quantum depletion $N_{\rm dep}$ is much smaller than the condensate fraction $N_{0}$. Here, the quantum depletion is expressed by 
\begin{eqnarray}
N_{\rm dep} = \sum_{m=-1}^{1}  \sum_{ \substack{k\neq 0 \\ k \in {\rm 1st BZ} } } v_{m,k}^2.
\end{eqnarray}
In our numerical calculation, $N_{\rm dep}/N_0$ is smaller than $0.014$. When the system size is too large, this condition breaks down. 

\subsection{Comparison between FQD and TWA results}
To demonstrate the validity of the TWA method, we compare FQD results with TWA ones.
Key parameters are $\kappa = \rho_{\rm f} J/U_0$ and the filling factor $\rho_{\rm f}$. When these parameters are much larger than unity, the TWA calculation becomes valid. Taking this condition into account, we apply the TWA method to scalar and spin-1 BH models. The former was investigated in Ref.~\cite{Pol_S}. 

\begin{figure}[t]
\begin{center}
\includegraphics[keepaspectratio, width=17.5cm,clip]{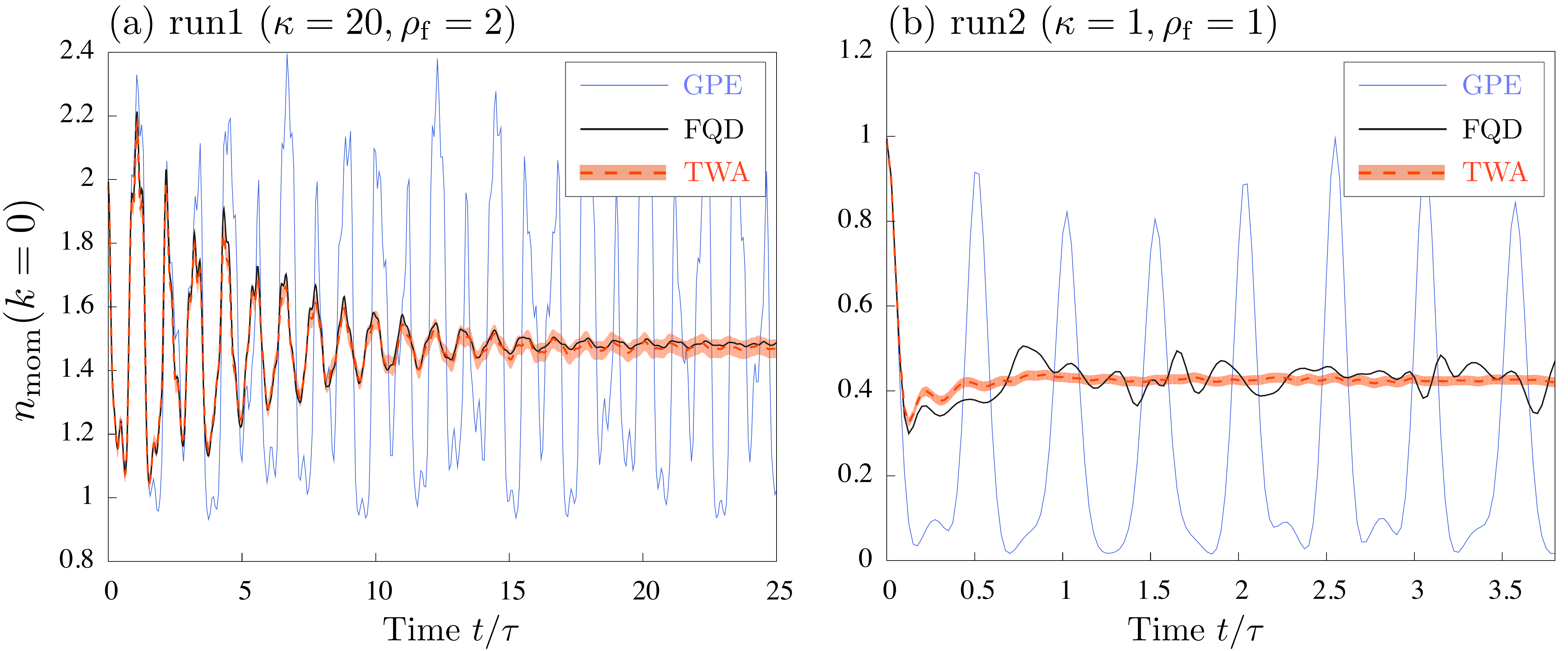}
\caption{Numerical results of the scalar BH model obtained by  FQD (black solid line), TWA (red dashed line), and the Gross-Pitaevskii equation (GPE) (blue solid line). These graphs show the time evolutions of the occupation number at zero momentum for (a) the run1 ($J/U_0=10$, $\kappa=20$, $\rho_{\rm f}=2$) and (b) the run2 ($J/U_0=1$, $\kappa=1$, $\rho_{\rm f}=1$). The abscissa is normalized by the characteristic time $\tau=4\hbar/J$. Color bands show $3\sigma/\sqrt{N_{\rm sam}}$ error bars in the TWA calculation, where $\sigma$ is the standard deviation. 
\label{scalar_ED}} 
\end{center}
\end{figure}

\subsubsection{Dynamics of the scalar BH model}
We perform FQD and TWA calculations in the scalar BH model defined by
\begin{eqnarray}
\hat{H}_{\rm scalar} = -J \sum_{j} \Bigl(\hat{b}_{j+1}^{\dagger} \hat{b}_{j} + {\rm h.c.} \Bigl) + \frac{U_0}{2} \sum_{j} \hat{n}_j(\hat{n}_j-1),  
\end{eqnarray}
where the parameters $J$ and $U_0$ are positive. We assume that the system has 5 lattices, and consider two parameter sets, namely, (a) the run1 ($J/U_0=10$, $\kappa=20$, $\rho_{\rm f}=2$) and (b) the run2 ($J/U_0=1$,  $\kappa=1$, $\rho_{\rm f}=1$). For the FQD calculations, we use the exact diagonalization method. Initial states in both (a) and (b) are coherent states where all particles occupy a single site. 

In this setup, we numerically compute the time evolution of the particle number $n_{\rm mom}(k=0)$ in the zero momentum state, which is defined by 
\begin{eqnarray}
n_{\rm mom}(k=0) = \frac{1}{M}\sum _{i,j} \langle \hat{b}_i^{\dagger} \hat{b}_j \rangle.
\end{eqnarray}
As shown in Fig.~\ref{scalar_ED}, we find that in the run (a) the TWA result agrees very well with the FQD one. On the other hand, in the run (b), the TWA result deviates slightly from the FQD result. In both cases, we also plot the mean field results obtained by the Gross-Pitaevskii equation (GPE) with an initial state $b_{j}=\sqrt{5\rho_{\rm f}} \delta_{j,1}$, which show large deviations from the FQD results. Thus, the TWA calculation becomes a good approximation when the conditions $\kappa \gg 1$ and $\rho_{\rm f} \gg 1$ are satisfied. 
 
\subsubsection{Dynamics of the spin-1 BH model}
The validity of the TWA method for the spin-1 BH model is discussed by comparing FQD calculations with TWA ones.
We consider a Hamiltonian given by
\begin{eqnarray}
\hat{H}_{\rm spin} =  -J \sum_{m,j} \Bigl(\hat{b}_{m,j+1}^{\dagger} \hat{b}_{m,j} + {\rm h.c.} \Bigl) + q \sum_{m,j} m^2 \hat{b}_{m,j}^{\dagger} \hat{b}_{m,j}  + \frac{U_0}{2} \sum_{j} \hat{n}_i(\hat{n}_j-1) + \frac{U_2}{2} \sum_j \Bigl( \hat{\bm{S}}_j^2 - 2 \hat{n}_j \Bigl). \label{BH}
\end{eqnarray}
We assume that the system is comprised of 3 lattices with the total particle number $N$ up to $24$ and subject to the periodic boundary condition; other parameters are set to be $J/U_0=20$ and $U_0/U_2=-1$ with $J>0$. In this model, the dimension of the Hilbert space is sufficinetly large, so that we solve the Schr$\rm \ddot{o}$dinger equation by using the Crank-Nicolson method \cite{CN} instead of the exact diagonalization. Similar to the scalar BH model, we investigate the time evolution of $n_{{\rm mom}, m}(k=0)~(m=1,0,-1)$ defined by 
\begin{eqnarray}
n_{{\rm mom}, m}(k=0) = \frac{1}{M} \sum _{i,j} \langle \hat{b}_{m,i}^{\dagger} \hat{b}_{m,j} \rangle, 
\end{eqnarray}
which is the occupation number of the zero-momentum state for the $m$-component. As shown in the previous numerical results in Fig.~\ref{scalar_ED}, the GP results deviate greatly from the FQD results in the spin-1 BH model as well. In the following figures, we omit the GP results.

\begin{figure}[t]
\begin{center}
\includegraphics[keepaspectratio, width=17.5cm,clip]{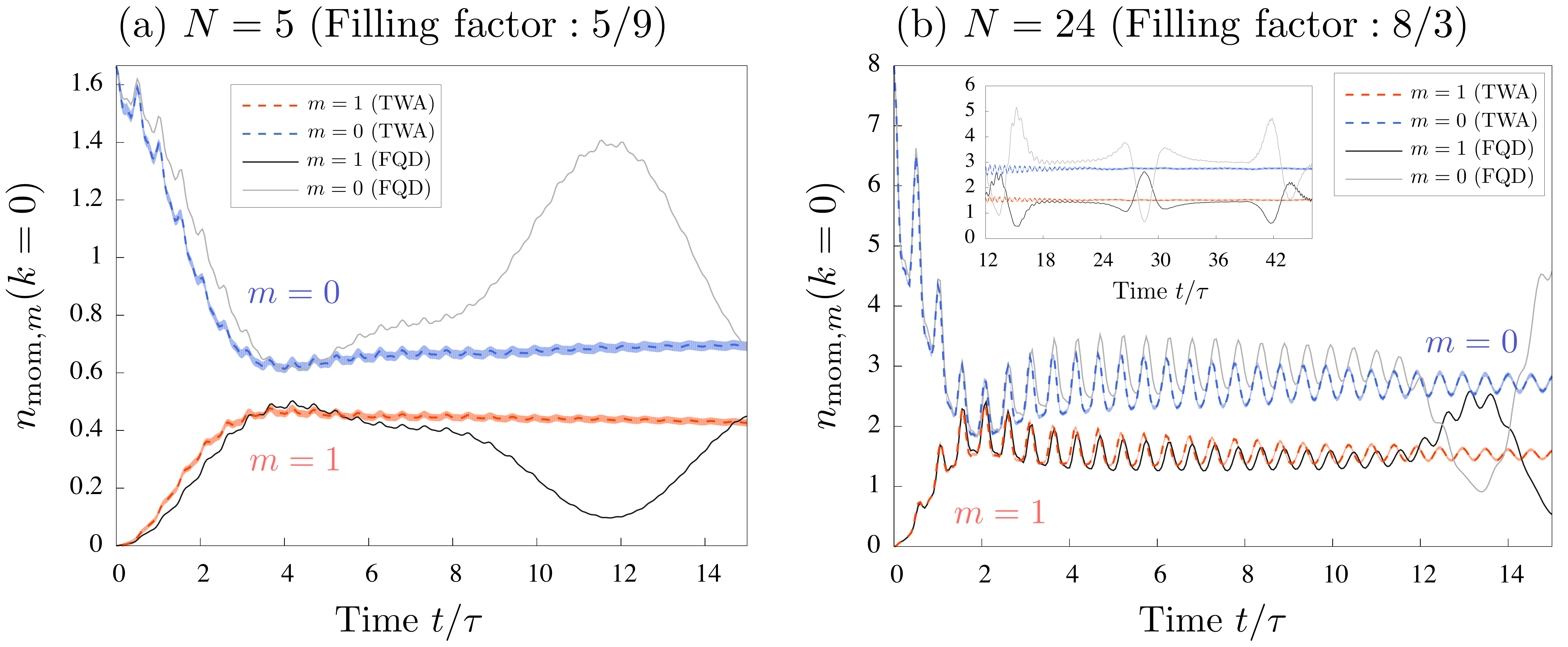}
\caption{Numerical results of the spin-1 BH model obtained by FQD and TWA without the quadratic Zeeman term. Figures (a) and (b) show the time evolutions of the occupation number at zero momentum for the total particle number $N=5$ and $N=24$, respectively. The inset of (b) shows numerical data for a long-time span ($12 \leq t/\tau \leq 46$). Color bands show $3\sigma/\sqrt{N_{\rm sam}}$ error bars in the TWA calculation.  \label{spinor_FQD1}} 
\end{center}
\end{figure}

Firstly, we show numerical results without the quadratic Zeeman term ($q=0$). The initial state in the TWA calculations is a coherent state where all particles occupy the $m=0$ component at a single site, but in the FQD calculations we use a Fock state corresponding to the TWA initial state. When the particle number $N$ is large, the difference between the coherent and Fock state is small. Thus, we can expect good agreement between the TWA and the FQD results for large $N$.

Figure \ref{spinor_FQD1} shows that the agreement between the TWA and FQD results dramatically improves as the total particle number increases. We also find that the excellent agreement between the FQD and TWA  results in  Fig.~\ref{spinor_FQD1} (b) breaks down when the first quantum revival-like behavior occurs ($t/\tau\sim13$). As shown in the inset of Fig.~\ref{spinor_FQD1} (b), this bump appears with a period $T/ \tau \sim 11$, so that we expect this revival-like behavior is caused by the smallness of the Hilbert space. However, the revival time $T$ grows exponentially large with increasing the number of particles. Experimentally, the quantum revival-like behavior is not usually observed when the particle number and the lattice number are of the order of $10,000$ and $100$, respectively. This implies that the period $T$ of the quantum revival-like behavior becomes quite long in comparison with a time scale ($\sim10~{\rm s}$) that actual experiments can access. As a related analytical calculation, the quantum revival in the Kerr model was investigated, and the period of the revival was proven to become very long when the particle number is large \cite{Pol_S}. From these results, we expect that the TWA method can be utilized to study the long-time dynamics such as coarsening dynamics when we study a large system ($N \gg 1$ and $M \gg 1$) in the deep superfluid regime ($\kappa \gg 1$).

\begin{figure}[t]
\begin{center}
\includegraphics[keepaspectratio, width=18cm,clip]{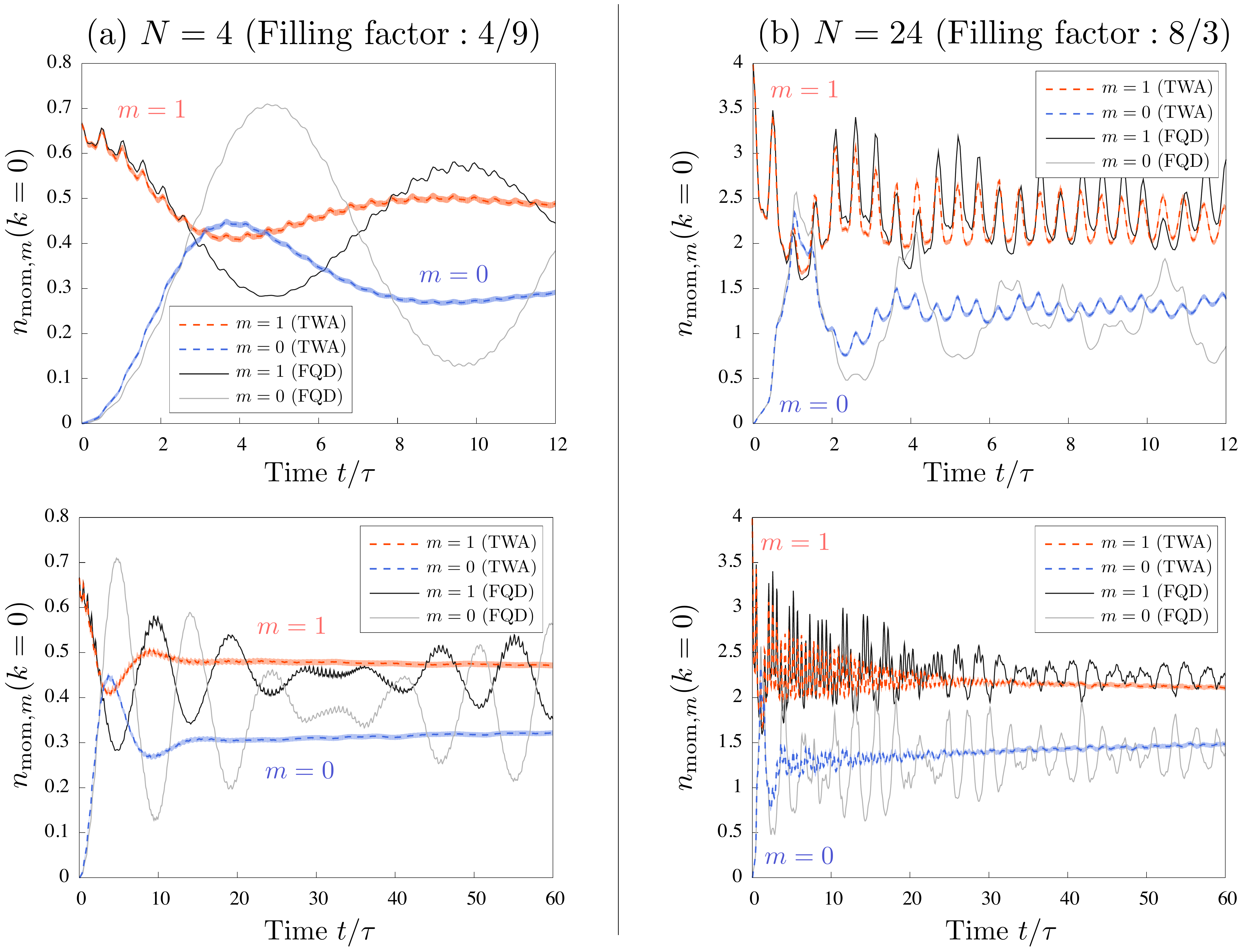}
\caption{Numerical results of the spin-1 BH model obtained by FQD and TWA in the presence of the quadratic Zeeman term. Figures (a) and (b) show the time evolutions of the occupation number at zero momentum for the total particle number $N=4$ and $N=24$, respectively. The upper and lower panels show the dynamics for the short-time ($0<t/\tau< 12$) and long-time ($0<t/\tau< 60$) dynamics. Color bands show $3\sigma/\sqrt{N_{\rm sam}}$ error bars in the TWA calculation.  \label{spinor_FQD2}} 
\end{center}
\end{figure}

Secondly, numerical results with the quadratic Zeeman term ($q=U_2N/M$) are discussed in Fig.~\ref{spinor_FQD2}.
The initial state in the TWA calculations is a coherent state where all particles equally occupy the $m=\pm 1$ components at a single site, and the initial state in the FQD calculations is a Fock state corresponding to the TWA initial state. When the total particle number is small, the agreement between the FQD and TWA results is not so good. However, the agreement becomes much better for a larger total particle number. For example, in the case with $N=4$, the TWA result for the $m=1$ component cannot follow the oscillation of the FQD result, but in the case with $N=24$ the TWA result well reproduces this oscillation. Thus, we can expect better agreement in the system where we study the coarsening dynamics because the total particle number is large. 

Here, by comparing between the results with and without the quadratic Zeeman term, we find that the deviation in the presence of the quadratic Zeeman term is large. We speculate that this is caused by the smallness of the Hilbert space. Despite the deviation, the time-averaged value for the FQD calculation is close to one for the TWA calculation. Furthermore, the main cause of the deviation of the $m=1$ component is the dynamics of the $m=0$ component. This $m=0$ component has a low occupation number, and is thus strongly affected by the quantum fluctuation. As a result, the $m=0$ mode leads to the deviation of the $m=1$ component because of the conservation of the total particle number. However, in our coarsening dynamics discussed in the main text, the particle number of the $m=1$ component is much larger than that of the $m=0$ component. Thus, the dynamics of the $m=1$ component is not affected by that of the $m=0$ component, and the TWA result in this case is expected to exhibit better agreement with the FQD one.
\\
 
\section{Analysis of a spin domain-wall pair in a one-dimensional Landau-Lifshitz equation}
To derive the growth law of the correlation length in coarsening dynamics of the 1D spin-1 BH model, it is necessary to investigate the dynamics of a spin domain-wall pair described by the 1D LL equation.  Firstly, we derive the 1D LL equation from Eq.~(\ref{CE}) by using three approximations. Secondly, we apply the singular perturbation method \cite{Kuramoto_S,KO_S,Ohta_S} to the LL equation, derive analytic expressions of this dynamics. Thirdly, performing numerical calculations, we find excellent agreement between the analytical and numerical results.

\subsection{Derivation of the Landau-Lifshitz equation} 
We can derive the Landau-Lifshitz equation from Eq.~(\ref{CE}) under the following three assumptions: (i) the  width $d$ of the domain wall is larger than the lattice constant $a$, (ii) the spin state is ferromagnetic, and (iii) the total density $\rho_{j}$ is independent of time $t$ and lattice site $j$.  As shown in the next section, we can change the width $d$ by controlling $J$ and $q$, so that the assumption (i) can be justified. Actually, in our numerical calculation, the ratio $d/a$ is about 6.4. As for the assumptions (ii) and (iii), we can also justify them by noting $U_{2}<0$ and $U_0 \gg |U_2|$ because the former condition ensures that the spin interaction is ferromagnetic and the latter condition suppresses the density fluctuation. 

The assumption (i) enables us to apply continuum approximation to Eq.~(\ref{CE}). Defining a macroscopic wave function as $\psi_m(x=ja) = b_{m,j}$, we obtain the spinor GP equation given by
\begin{eqnarray}
i \frac{\partial }{\partial t} \psi_{m}(x,t) = -J'\frac{\partial^2}{\partial x^2} \psi_{m}(x,t) - q' m^2 \psi_{m}(x,t) + U_{0}'  \rho (x,t) \psi_{m}(x,t) + U_{2}'\sum_{n=-1}^1  (\bm{S}(x,t) \cdot \hat{\bm{S}})_{mn}  \psi_{n}(x,t), \label{SpinorGP}
\end{eqnarray}
where $J'=Ja^2/\hbar$, $q'=-q_{\rm F}/\hbar$, $U_{0}'=U_{0}/\hbar$, $U_{2}'=U_{2}/\hbar$, $\rho (x,t) = \sum_{m=-1}^1 |\psi_{m}(x,t)|^2$, $\bm{S} (x,t) = \sum_{n,m=-1}^1 \psi_{n}(x,t)^* (\hat{\bm{S}})_{nm} \psi_{m}(x,t)$. Here, we omit a chemical potential term because it can be removed by a $U(1)$ transformation. Using the assumption (ii) and Eq.~(\ref{SpinorGP}), we can derive the spin hydrodynamics equation \cite{YU} given by 
\begin{eqnarray}
\frac{\partial}{\partial t} \rho(x,t) +  \frac{\partial }{\partial  x} \biggl( \rho(x,t) v(x,t) \biggl) = 0, \label{SH1}
\end{eqnarray}
\begin{eqnarray}
\frac{\partial}{\partial t} S_{\mu}(x,t) +  \frac{\partial }{\partial  x} J_{\mu}(x,t) = \frac{q'}{\rho(x,t)} S_z(x,t)(  \bm{S}(x,t) \times \hat{\bm{e}}_z )_{\mu}, \label{SH2}
\end{eqnarray}
\begin{eqnarray}
J_{\mu}(x,t) =  S_{\mu}(x,t) v(x,t) - J' \sum _{\nu,\lambda=x,y,z} \epsilon _{\mu \nu \lambda}  S_{\nu}(x,t) \frac{\partial }{\partial  x} \Biggl( \frac{S_{\lambda}(x,t)}{\rho (x,t)} \Biggl), \label{SH3}
\end{eqnarray}
\begin{eqnarray}
\frac{\partial}{\partial t}  v(x,t)&+& v(x,t) \frac{\partial}{\partial x} v(x,t) - 2J'^{2} \frac{\partial}{\partial x} \Biggl(  \frac{1}{\rho(x,t)} \frac{\partial^2}{\partial x^2} \sqrt{\rho(x,t)} \Biggl)  \nonumber  \\ 
&+& \frac{J'^{2}}{\rho (x,t)} \frac{\partial }{\partial  x}  \sum_{\nu=x,y,z}\Biggl\{  \rho(x,t) \biggl( \frac{\partial }{\partial  x} \frac{S_{\nu}(x,t)}{\rho(x,t)} \biggl)^2
- S_{\nu}(x,t)   \frac{\partial^2 }{\partial  x^2} \frac{S_{\nu}(x,t)}{\rho(x,t)}  \Biggr\}  \nonumber  \\ 
&=& -2J' \biggl\{ U_{0}' \frac{\partial}{\partial x} \rho(x,t) + U_{2}' \sum_{\nu=x,y,z}\frac{S_{\nu}(x,t)}{\rho(x,t)} \frac{\partial}{\partial x} S_{\nu}(x,t) \biggr\},  \label{SH4}
\end{eqnarray}
where $v(x,t)$ is the velocity. Finally, we make use of the assumption (iii) in Eqs.~(\ref{SH1})--(\ref{SH4}) and denote the uniform density as $\rho(x,t) = \rho_0$. 
We then obtain the following equations: 
\begin{eqnarray}
\frac{\partial }{\partial  x} v(x,t)  = 0, \label{SH5}
\end{eqnarray}
\begin{eqnarray}
\frac{\partial}{\partial t} S_{\mu}(x,t) +  v(x,t) \frac{\partial }{\partial  x} S_{\mu}(x,t)
= \frac{J'}{\rho_0} \sum _{\nu,\lambda=x,y,z} \epsilon _{\mu \nu \lambda}  S_{\nu}(x,t) \frac{\partial^2 }{\partial  x^2} S_{\lambda}(x,t)
+ \frac{q'}{\rho_0} S_z(x,t)(  \bm{S}(x,t) \times \hat{\bm{e}}_z )_{\mu}, \label{SH6}
\end{eqnarray}
\begin{eqnarray}
\frac{\partial}{\partial t}  v(x,t) + \frac{J'^{2}}{\rho_0^2} \frac{\partial }{\partial  x}  \sum_{\nu=x,y,z}\Biggl\{   \biggl( \frac{\partial }{\partial  x} S_{\nu}(x,t) \biggl)^2
- S_{\nu}(x,t)   \frac{\partial^2 }{\partial  x^2} S_{\nu}(x,t)  \Biggr\}  = 0, \label{SH7}
\end{eqnarray}
where in the derivation of Eqs.~(\ref{SH6}) and (\ref{SH7}) we use Eq.~(\ref{SH5}) and $\bm{S}(x,t)^2 = \rho(x,t)^2$ which follows from the assumption (ii). Equation~(\ref{SH5}) shows that the velocity is independent of space $x$, so that we can write $v(x,t)=v(t)$. Substituting this expression into Eq.~(\ref{SH7}) and integrating it in the $x$-direction, we can obtain $d v(t) /d t =0$ by integrating the third term in Eq.~(\ref{SH7}). Thus, by denoting  this uniform velocity as $v_0$, Eq.~(\ref{SH6}) becomes   
\begin{eqnarray}
\frac{\partial}{\partial t} S_{\mu}(x,t) +  v_0 \frac{\partial }{\partial  x} S_{\mu}(x,t)
= \frac{J'}{\rho_0} \sum _{\nu,\lambda=x,y,z} \epsilon _{\mu \nu \lambda}  S_{\nu}(x,t) \frac{\partial^2 }{\partial  x^2} S_{\lambda}(x,t)
+ \frac{q'}{\rho_0} S_z(x,t)(  \bm{S}(x,t) \times \hat{\bm{e}}_z )_{\mu}. \label{SH8}
\end{eqnarray}
The second term on the left-hand side of Eq.~(\ref{SH8}) just induces a parallel translation, which can be removed by a Galilean transformation.
Thus, this inertial term is not important when we consider the domain-wall dynamics dominated by a genuine domain-wall interaction. 
As a result, after the transformation, we obtain the LL equation given by
\begin{eqnarray}
\rho_0 \frac{\partial}{\partial t} {\bm S}(x,t) = -\bm{S}(x,t) \times \bm{B}(x,t),  \label{LL1}
\end{eqnarray}
\begin{eqnarray}
\bm{B}(x,t) = -J'\nabla^2 \bm{S}(x,t) - q' S_z\hat{\bm e}_z.  \label{LL2}
\end{eqnarray}
Here, the spin vector satisfies a relation $|\bm{S}(x,t)| = \rho_0$, and we can eliminate $\rho_0$ in Eqs.~(\ref{LL1}) and (\ref{LL2}) by replacing $\bm{S}$ by $\rho_0 \bm{S}$.
In all of the following calculations, we use Eqs.~(\ref{LL1}) and (\ref{LL2}) after this elimination.
\\

\subsection{A domain wall solution of the Landau-Lifshitz equation}
The LL equation (\ref{LL1}) and (\ref{LL2}) has a stationary domain-wall solution given by 
\begin{eqnarray}
S_x^{\rm d}(x) = {\rm cos}\phi_{\rm d} ~{\rm sech}(x/\lambda),  \label{LL3}
\end{eqnarray}
\begin{eqnarray}
S_y^{\rm d}(x) = {\rm sin}\phi_{\rm d} ~{\rm sech}(x/\lambda), \label{LL4}
\end{eqnarray}
\begin{eqnarray}
S_z^{\rm d}(x) = {\rm tanh}(x/\lambda),  \label{LL5}
\end{eqnarray}
where the azimuthal angle $\phi_{\rm d}$ is an arbitrary constant and $\lambda = \sqrt{J'/q'}=d/2$ is the half-width of the wall.
This solution is obtained by solving $\bm{S}(x) \times \bm{B}(x)=0$ with the boundary condition $S_z(x) \rightarrow \pm 1 \hspace{0.5mm} (x \rightarrow \pm \infty)$ \cite{Landau,Tatara_S}.

\subsection{Stereographic projection for the spin vector}
\begin{figure}[t]
\begin{center}
\includegraphics[keepaspectratio, width=15cm,clip]{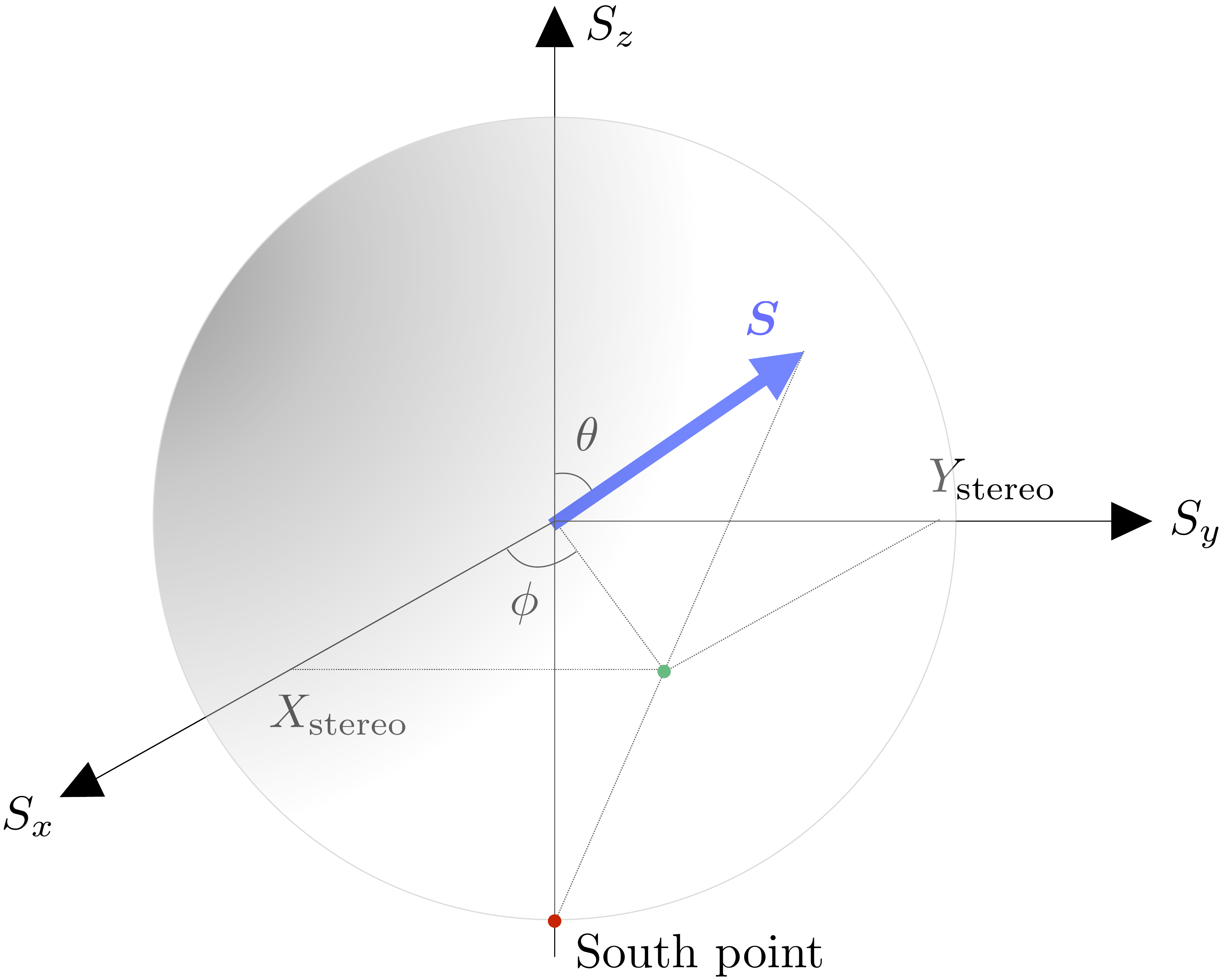}
\caption{Stereographic projection for the spin vector $\bm{S}$. Angles $\phi$ and $\theta$ are the azimuthal and polar angles of $\bm{S}$, respectively. In this figure, the projection is performed from the south pole. Two variables $X_{\rm stereo}$ and $Y_{\rm stereo}$ represent the coordinates in the $S_x$-$S_y$ plane.
\label{stereo}} 
\end{center}
\end{figure}

In the following calculation, it is convenient to use the stereographic projection \cite{Nakamura_S} for the spin vector $\bm{S}(x,t)$ as shown in Fig.~\ref{stereo}. This transformation introduces a new complex variable defined by 
 \begin{eqnarray}
\eta(x,t) &=& X_{\rm stereo}(x,t) + i Y_{\rm stereo}(x,t) \nonumber \\
&=& e^{i\phi(x,t)}{\rm tan} \Big( \theta(x,t)/2 \Big),  \label{LL6}
\end{eqnarray}
with the azimuthal angle $\phi(x,t)$ and the polar angle $\theta(x,t)$ for the spin vector $\bm{S}(x,t)$. 
Substituting Eq.~\eqref{LL6} into Eq.~\eqref{LL1}, we derive the equation of motion for $\eta(x,t)$:
\begin{eqnarray}
i \frac{\partial}{\partial t} \eta(x,t) = -\frac{\partial^2}{\partial x^2}\eta(x,t) + F(\eta(x,t)),  \label{LL7}
\end{eqnarray}
\begin{eqnarray}
F(\eta(x,t)) = \frac{2\eta(x,t)^*}{1+|\eta(x,t)|^2} \biggl( \frac{\partial}{\partial x}\eta(x,t) \biggl)^2 +  \frac{1-|\eta(x,t)|^2}{1+|\eta(x,t)|^2}\eta(x,t).  \label{LL7_1}
\end{eqnarray}
Here, for the simplicity of the notation, $J'$ and $q'$ are set to be unity. 
In the following, we apply the singular perturbation method \cite{Kuramoto_S,KO_S,Ohta_S} to Eq.~\eqref{LL7}. The reason we use this method is as follows. If we apply a usual perturbation method to the LL equation, a perturbation expansion is found to break down because of the existence of the zero mode (uniform spatial translation of a domain wall). This mode causes divergence of the usual perturbative solution. This is often called a secular problem, and it is well known that we can remove such a divergence by using the singular perturbation method \cite{Kuramoto_S,KO_S,Ohta_S}.

\begin{figure}[t]
\begin{center}
\includegraphics[keepaspectratio, width=16.5cm,clip]{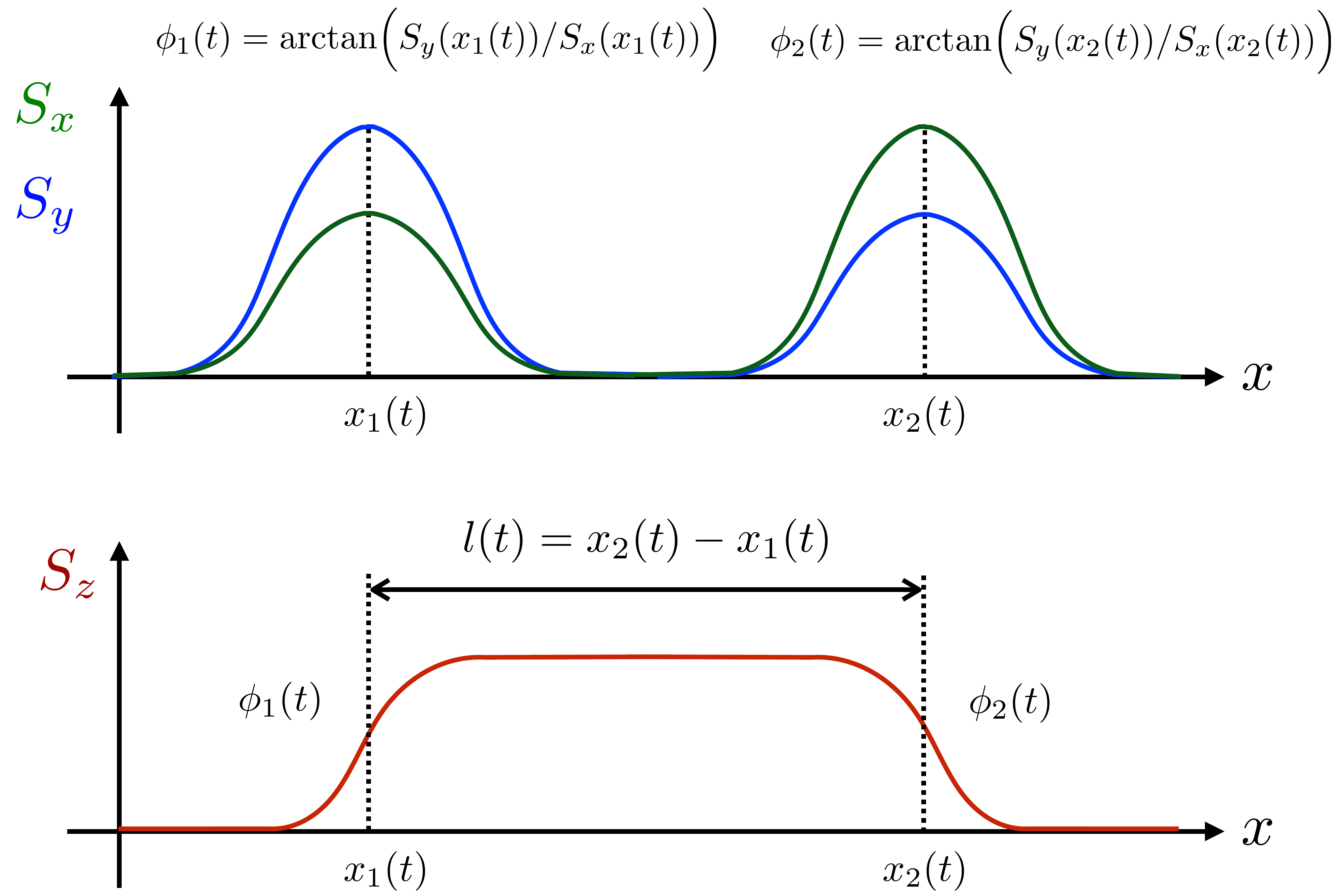}
\caption{Schematic spin configuration $\bm{S}(x,t)$ for a spin-domain pair. Time-dependent variables $x_j(t)~(j=1,2)$ and $\phi_j(t)~(j=1,2)$ are the positions and the azimuthal angles at the center of domain walls, respectively. Here $l(t)=x_2(t)-x_1(t)$ denotes the distance between the walls. 
\label{LL_spin_configuration}} 
\end{center}
\end{figure}

\subsection{Spin configuration for the domain pair}
The situation we consider is depicted in Fig.~\ref{LL_spin_configuration}, where two spin domains are spaced by distance $l(t)$. Positions of the walls are specified by time-dependent variables $x_1(t)$ and $x_2(t)$, and the distance is $l(t)=x_2(t)-x_1(t)$. The distinction between the previous work \cite{KO_S,Ohta_S} and ours is the existence of $\phi_1(t)$ and $\phi_2(t)$, which are the azimuth angles of the spin vector at the center of domain walls. Using the single domain wall solution of Eq.~(\ref{LL3})-(\ref{LL5}), we can approximately express the spin configuration in Fig.~\ref{LL_spin_configuration} as 
\begin{eqnarray}
S_x^{\rm dp}(x) = \frac{1}{S^{\rm dp}(x)}\biggl[ S_x^{\rm d}(X_1,\phi_1) + S_x^{\rm d}(X_2,\phi_2) \biggl],   \label{LL8}
\end{eqnarray}
\begin{eqnarray}
S_y^{\rm dp}(x) = \frac{1}{S^{\rm dp}(x)}\biggl[ S_y^{\rm d}(X_1,\phi_1) + S_y^{\rm d}(X_2,\phi_2) \biggl],   \label{LL9}
\end{eqnarray}
\begin{eqnarray}
S_z^{\rm dp}(x) = \frac{1}{S^{\rm dp}(x)}\biggl[ S_z^{\rm d}(X_1) - S_z^{\rm d}(X_2) -1 \biggl],  \label{LL10}
\end{eqnarray}
\begin{eqnarray}
S^{\rm dp}(x) = \sqrt{ \Bigl[S_x^{\rm d}(X_1,\phi_1) + S_x^{\rm d}(X_2,\phi_2) \Bigl]^2 +  \Bigl[S_y^{\rm d}(X_1,\phi_1) + S_y^{\rm d}(X_2,\phi_2) \Bigl]^2  
+  \Bigl[ S_z^{\rm d}(X_1) - S_z^{\rm d}(X_2) -1 \Bigl]^2  }, 
\label{LL10_1}
\end{eqnarray}
with $X_{j} = x-x_j(t) ~(j=1,2)$. Then, a stereographic variable $\eta_{\rm dp}(x)$ for the domain-wall pair is represented as 
\begin{eqnarray}
\eta_{\rm dp}(x) = {\rm exp}(i\phi_{\rm dp}(x)) {\rm tan}\Biggl( \frac{\theta_{\rm dp}(x)}{2} \Biggl),   \label{LL11}
\end{eqnarray}
\begin{eqnarray}
{\rm cos}\bigl(\theta_{\rm dp}(x) \bigl) = S_{z}^{\rm dp}(x),   \label{LL12}
\end{eqnarray}
\begin{eqnarray}
{\rm tan}\bigl(\phi_{\rm dp}(x) \bigl) &=& \frac{S_{y}^{\rm dp}(x)}{S_{x}^{\rm dp}(x)}.   \label{LL13}
\end{eqnarray}

\subsection{Application of the singular perturbation method to Eq.~(\ref{LL7})}
In the following analysis, we regard ${\rm exp}(-l(t)/\lambda)$ as a small quantity. We make three assumptions: (I) the distance $l(t)$ between the walls is much larger than the width of the wall $\lambda$, (II) the deformation $\rho_{\rm def}(x,t)$ from Eq.~(\ref{LL11}) is of the same order as ${\rm exp}(-l(t)/\lambda)$ (a definition of $\rho_{\rm def}(x,t)$ is given in Eq.~(\ref{LL21})), and (III) the velocity $dx_j(t)/dt~(j=1,2)$ and frequency $d\phi_{j}(t)/dt ~(j=1,2)$ of the domain walls are also of the same order as ${\rm exp}(-l(t)/\lambda)$. Under these assumptions, we focus on a region $|X_2(t)|=|x-x_2(t)| \lesssim \lambda$, and derive the equations of motion for $x_2(t)$ and $\phi_2(t)$. We note $\lambda=1$ because of $J'=1$ and $q'=1$ in our above notation.

To begin with, we derive an expression of $\eta(x,t)$ in the region $|X_2(t)| \lesssim 1$. Using the approximation (I), we keep terms up to the first order in ${\rm exp}(-l)$ and obtain
\begin{eqnarray}
{\rm sech}(X_1) &=& {\rm sech}(X_2+l) \nonumber \\ 
&=& \frac{2{\rm exp}(-X_2-l)}{1+{\rm exp}(-2X_2-2l)}  \nonumber \\
&\simeq& 2{\rm exp}(-X_2-l),   \label{LL14}
\end{eqnarray}
\begin{eqnarray}
{\rm tanh}(X_1) &=& {\rm tanh}(X_2+l) \nonumber \\ 
&=& \frac{1-{\rm exp}(-2X_2-2l)}{1+{\rm exp}(-2X_2-2l)}  \nonumber \\
&\simeq& 1 - 2{\rm exp}(-2X_2-2l) \nonumber \\
&\simeq& 1.   \label{LL15}
\end{eqnarray} 
Then, these expansions lead to 
\begin{eqnarray}
{\rm tan}\bigl(\phi_{\rm dp}(x) \bigl) &=& \frac{S_{y}^{\rm dp}(x)}{S_{x}^{\rm dp}(x)} \nonumber \\
&=& \frac{S_y^{\rm d}(X_1,\phi_1) + S_y^{\rm d}(X_2,\phi_2)}{S_x^{\rm d}(X_1,\phi_1) + S_x^{\rm d}(X_2,\phi_2)} \nonumber \\
&=& \frac{ 2{\rm exp}(-X_2-l)~{\rm sin}\phi_1 + {\rm sin}\phi_2 ~{\rm sech}(X_2) }{2{\rm exp}(-X_2-l)~{\rm cos}\phi_1 + {\rm cos}\phi_2 ~{\rm sech}(X_2)}  \nonumber \\
&\simeq& {\rm tan}(\phi_2) + \frac{2{\rm exp}(-X_2-l)~{\rm sin}\phi_1 }{{\rm cos}\phi_2 ~{\rm sech}(X_2)}  - \frac{2{\rm exp}(-X_2-l)~{\rm cos}\phi_1{\rm sin}\phi_2}{ ({\rm cos}\phi_2)^2 ~{\rm sech}(X_2)} \nonumber \\
&=& {\rm tan}(\phi_2) + \frac{2{\rm exp}(-X_2-l)~{\rm sin}(\phi_1-\phi_2)}{({\rm cos}\phi_2)^2{\rm sech}(X_2)} \nonumber \\
&\simeq& {\rm tan}\Biggl( \phi_2 + \frac{2{\rm exp}(-X_2-l)~{\rm sin}(\phi_1-\phi_2)}{{\rm sech}(X_2)}  \Biggl).
\label{LL16}
\end{eqnarray}
As a result, the azimuthal angle is given by
\begin{eqnarray}
\phi_{\rm dp}(x) = \phi_2 + H(X_2,\phi_1-\phi_2){\rm exp}(-l), \label{LL17}
\end{eqnarray}
\begin{eqnarray}
H(X_2,\phi_1-\phi_2) =  2{\rm sin}(\phi_1-\phi_2){\rm exp}(-X_2){\rm cosh}(X_2). \label{LL18}
\end{eqnarray}
A similar calculation leads to 
\begin{eqnarray}
{\rm tan}\Biggl( \frac{\theta_{\rm dp}(x)}{2} \Biggl) &=& \frac{\sqrt{ 1- \bigl( {\rm cos}\theta_{\rm dp}(x) \bigl)^2}  }{1+{\rm cos}\theta_{\rm dp}(x)} \nonumber \\ 
&=& \frac{\sqrt{1- \bigl( S_{z}^{\rm dp}(x) \bigl)^2}}{1+S_{z}^{\rm dp}(x)} \nonumber \\
&\simeq& {\rm exp}(X_2) \biggl[  1 + G(X_2,\phi_1-\phi_2){\rm exp}({-l}) \biggl], \label{LL19}
\end{eqnarray}
where the function $G(X_2,\phi_1-\phi_2)$ is given by 
\begin{eqnarray}
G(X_2,\phi_1-\phi_2) =  -2{\rm cos}(\phi_1-\phi_2){\rm exp}(-X_2){\rm sinh}(X_2). \label{LL18_1}
\end{eqnarray}
Thus, we substitute Eqs.~\eqref{LL17} and \eqref{LL19} into \eqref{LL11}, obtaining
\begin{eqnarray}
\eta_{\rm dp}(x) =  \biggl[  1 + G(X_2,\phi_1-\phi_2){\rm exp}({-l}) \biggl] {\rm exp}\biggl( X_2 + i\phi_2 + iH(X_2,\phi_1-\phi_2){\rm exp}(-l) \biggl).    \label{LL20}
\end{eqnarray}

Finally, to consider the deformation of a domain configuration, we introduce the function $\rho_{\rm def}(x,t)$ as follows:
\begin{eqnarray}
\eta(x,t) &=& {\rm exp} \biggl( X_2(t) + i\phi_2(t) + \Bigl[ G(X_2(t),\phi_1(t)-\phi_2(t))+iH(X_2(t),\phi_1(t)-\phi_2(t)) \Bigl] {\rm exp}(-l(t)) +2\rho_{\rm def}(x,t) {\rm cosh}X_2(t) \biggl) \nonumber \\
&\simeq& \eta_{\rm d}(x,t) \biggl( 1 + \epsilon(x,t) \biggl),     \label{LL21}
\end{eqnarray}
where
\begin{eqnarray}
\eta_{\rm d}(x,t) = {\rm exp} \biggl( X_2(t) + i\phi_2(t) \biggl),    \label{LL22}
\end{eqnarray}
\begin{eqnarray}
\epsilon(x,t) = \Bigl[ G(X_2(t),\phi_1(t)-\phi_2(t))+ iH(X_2(t),\phi_1(t)-\phi_2(t)) \Bigl] {\rm exp}(-l(t)) +2\rho_{\rm def}(x,t) {\rm cosh}X_2(t).   \label{LL22}
\end{eqnarray}
This is an approximate expression of $\eta(x,t)$ in the region $|X_2(t)| \lesssim 1$. 

Substituting Eqs.~\eqref{LL21}-\eqref{LL22} into Eq.~\eqref{LL7} and keeping terms up to the first order in the small quantities (assumptions (I)-(III)), 
we obtain an equation of motion for $\rho_{\rm def}(x,t)$:
\begin{eqnarray}
i\frac{\partial}{\partial T}\rho_{\rm def}(X_2,T)  = & & \mathcal{L} \rho_{\rm def}(X_2,T) \nonumber \\ 
& &+ \frac{ \dot{\phi}_2(T)+i \dot{x}_2(T)}{2{\rm cosh}(X_2)} +  \frac{ \Bigl[ G'(X_2,\phi_1(T)-\phi_2(T))+iH'(X_2,\phi_1(T)-\phi_2(T)) \Bigl]e^{X_2-l(T)}}{ {\rm cosh} (X_2)^2},   \label{LL23}
\end{eqnarray}
\begin{eqnarray}
\mathcal{L} = -\frac{\partial^2}{\partial X_2^2} + 1 -\frac{2}{ ({\rm cosh}X_2)^2}, \label{LL24}
\end{eqnarray}
where we perform the transformations $X_2(t) = x - x_2(t)$ and $T=t$, and use notations $H'=dH/dx$ and $G'=dG/dx$.

Here, we consider an eigenvalue problem defined by 
\begin{eqnarray}
\mathcal{L} \Psi_n(X_2) = E_n \Psi_n(X_2), 
\end{eqnarray}
which is known to have a zero mode \cite{Tatara_S,Landau_QM}. The eigenfunction for the zero mode is given by
\begin{eqnarray}
\Psi_0(X_2) = \frac{1}{{\rm cosh}(X_2)} \label{LL25}.
\end{eqnarray}

Thus, as in Refs. \cite{Kuramoto_S,KO_S,Ohta_S}, a solvable condition becomes
\begin{eqnarray}
\int_{-\lambda}^{{\lambda}}  \Psi_0(X_2)^* \rho_{\rm def}(X_2,T) dX_2 \simeq  \int_{-\infty}^{{\infty}}  \Psi_0(X_2)^* \rho_{\rm def}(X_2,T) dX_2 = 0\label{LL26}, 
\end{eqnarray}
which removes the divergence of the solution of Eq.~(\ref{LL23}) caused by the zero mode (secular term).  
From Eqs.~\eqref{LL23} and \eqref{LL26}, we obtain
\begin{eqnarray}
- \frac{\partial}{\partial T}  \phi_2(T) - i \frac{\partial}{\partial T} x_2 (T) &=& \int_{-\infty}^{\infty}  \frac{1}{{\rm cosh}(X_2) } \frac{ e^{-l(T)} \Bigl[ G'(X_2,\phi_1(T)-\phi_2(T))+iH'(X_2,\phi_1(T)-\phi_2(T)) \Bigl] e^{X_2}}{{\rm cosh}(X_2)^2}dX_2  \nonumber \\
&=&  -4i e^{-l(T)} {\rm sin}(\phi_1(T)-\phi_2(T)) -4 e^{-l(T)} {\rm cos}(\phi_1(T)-\phi_2(T)),  \label{LL27}
\end{eqnarray}
which leads to 
\begin{eqnarray}
\frac{\partial}{\partial T} x_2 (T) =  4 e^{-l(T)} {\rm sin}(\phi_1(T)-\phi_2(T)),  \label{LL28}
\end{eqnarray}
\begin{eqnarray}
\frac{\partial}{\partial T} \phi_2 (T)=4 e^{-l(T)} {\rm cos}(\phi_1(T)-\phi_2(T)).  \label{LL29}
\end{eqnarray}
As for an equation of $X_1$ and $\phi_1$, we can similarly derive
\begin{eqnarray}
\frac{\partial}{\partial T}x_1 (T) =  4 e^{-l(T)} {\rm sin}(\phi_1(T)-\phi_2(T)),  \label{LL30}
\end{eqnarray}
\begin{eqnarray}
\frac{\partial}{\partial T}\phi_1 (T)=4 e^{-l(T)} {\rm cos}(\phi_1(T)-\phi_2(T)).  \label{LL31}
\end{eqnarray}
These results show that the domain pair undergoes a linear uniform motion at a constant velocity $V(l)$ while rotating $S_x$ and $S_y$ at a constant frequency $\Omega(l)$:
\begin{eqnarray}
V(l) =  4e^{-l} {\rm sin}(\phi_1-\phi_2),  \label{LL32}
\end{eqnarray}
\begin{eqnarray}
\Omega(l) =  4e^{-l} {\rm cos}(\phi_1-\phi_2),  \label{LL32_1}
\end{eqnarray}
where both the phase difference $\phi_1-\phi_2$ and the distance $l$ are constant.
In the notation of $J'$ and $q'$, Eqs.~(\ref{LL32}) and (\ref{LL32_1}) become 
\begin{eqnarray}
V(l) =  4 \sqrt{J' q'} e^{-l/\lambda} {\rm sin}(\phi_1-\phi_2),  \label{LL33}
\end{eqnarray}
\begin{eqnarray}
\Omega(l) =  4 q' e^{-l/\lambda} {\rm cos}(\phi_1-\phi_2).  \label{LL33_1}
\end{eqnarray}
As in the assumptions (I), (II), and (III), this analytic expression is valid when the following two conditions are satisfied: (i) the distance $l=x_2-x_1$ between the walls is much larger than the width of the wall $\lambda$, and (ii) the angle difference $\phi_1-\phi_2$ is far from an integer multiple of $\pi/2$. 

Equations (\ref{LL33}) and (\ref{LL33_1}) have two features. One is an exponential dependence on the distance $l$ between two walls, and the other is the dependence on $\phi_1-\phi_2$. The first feature reflects the interaction induced by an exponential tail of the domain wall configuration. The second one is related to the vector product in the LL equation (\ref{LL1}). This is because the direction of this torque term is determined by the directions of transverse spins of the two interacting domain walls. 

\begin{figure}[t]
\begin{center}
\includegraphics[keepaspectratio, width=18cm,clip]{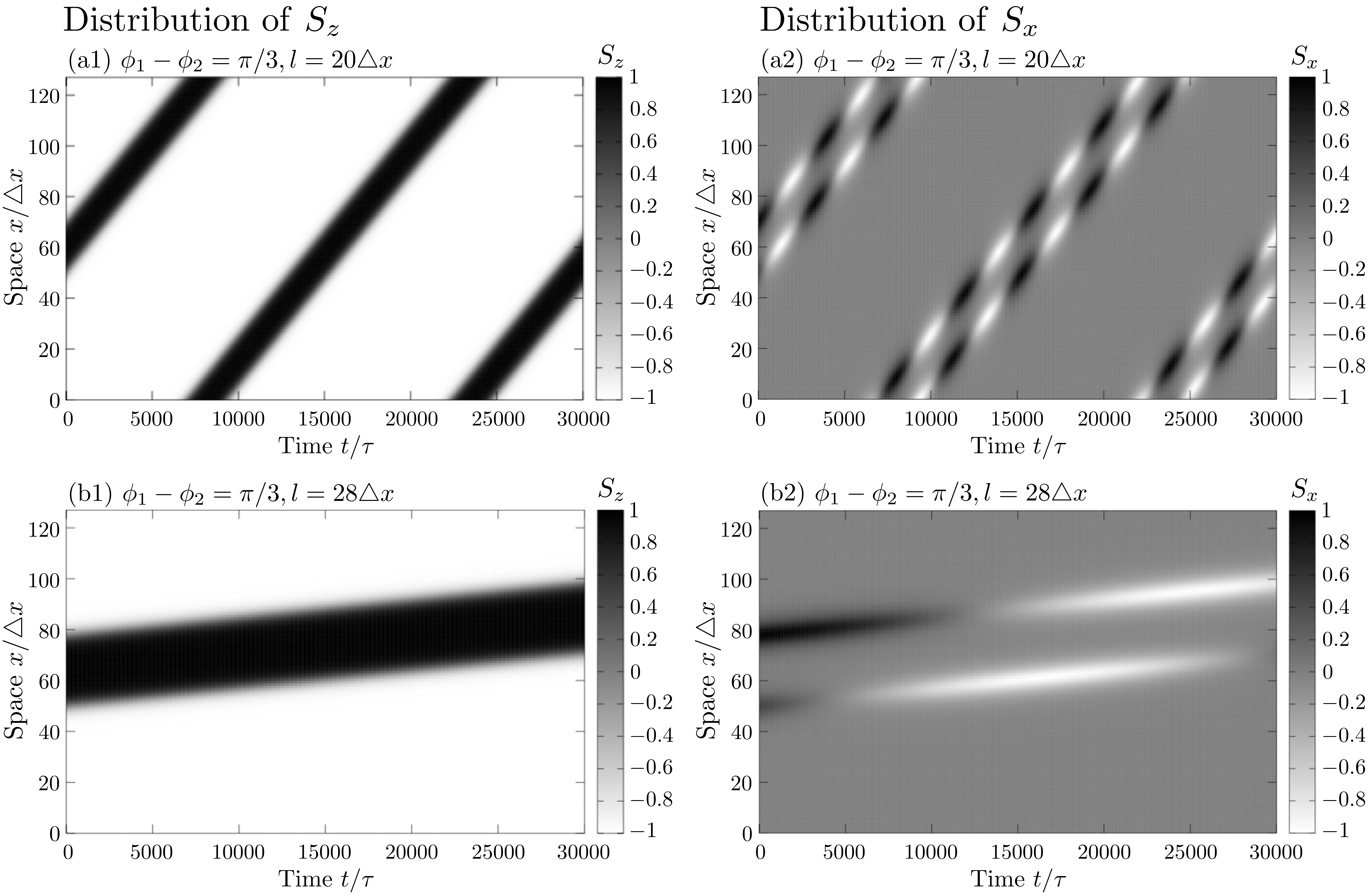}
\caption{Numerical results with the angle difference $\phi_1-\phi_2=\pi/3$ for the 1D LL equation. 
The left and right figures show the time evolutions of $S_z(x,t)$ and $S_x(x,t)$, respectively.
The space and time are normalized by the numerical mesh $\triangle x = \lambda/3.2$ and the characteristic time $\tau=4\hbar/J$.
\label{quarter_pi} }
\end{center}
\end{figure}

\subsection{Numerical test of the analytical result}
We numerically calculate the 1D LL equation \eqref{LL1} to investigate the validity of Eqs.~(\ref{LL32}) and (\ref{LL32_1}). Taking the conditions (i) and (ii) into account, we perform numerical calculations corresponding to $\phi_1-\phi_2 = \pi/12, \pi/6, \pi/4, \pi/3, 5\pi/12$ by varying the distance $l$.

\begin{figure}[t]
\begin{center}
\includegraphics[keepaspectratio, width=17.5cm,clip]{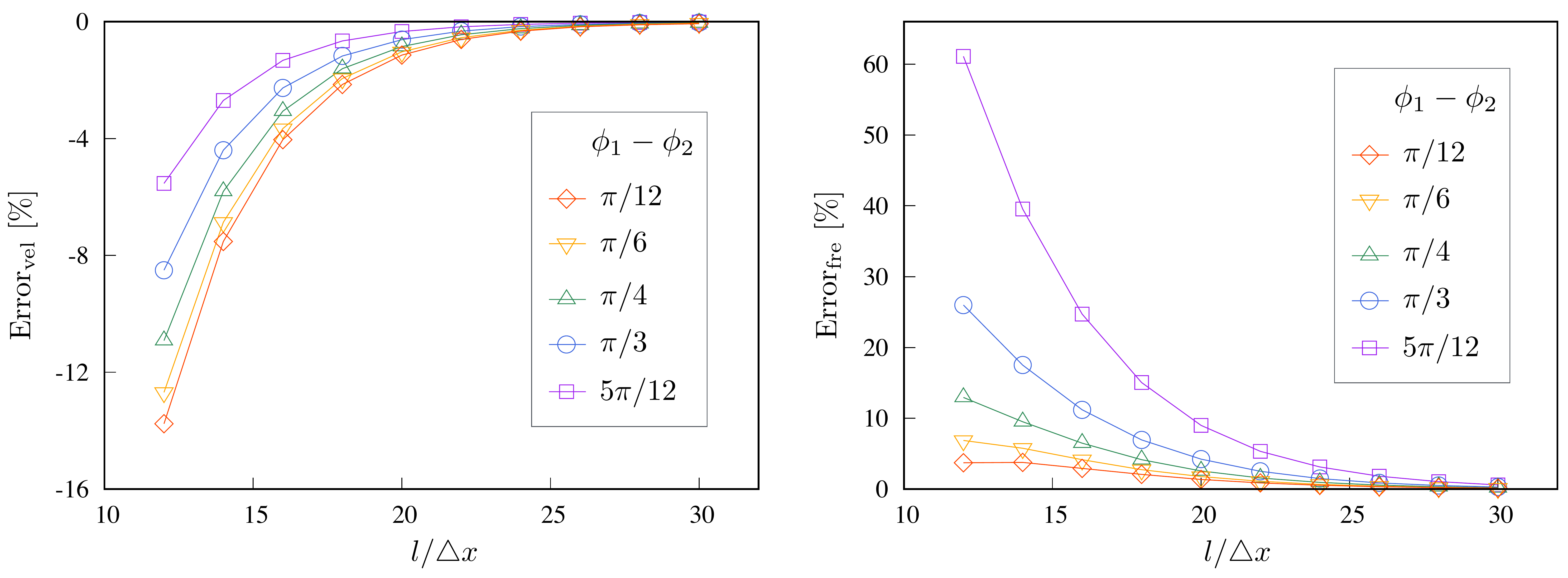}
\caption{Numerical results of the quantities ${\rm Error}_{\rm vel}$ (left) and ${\rm Error}_{\rm fre}$ (right) for the 1D LL equation. The analytical expressions of Eqs.~\eqref{LL32} and \eqref{LL32_1} show excellent agreement with the numerical results when the following two conditions are satisfied: (i) the distance $l=x_2-x_1$ between the domain walls is much larger than the width of the wall $\lambda$, and (ii) the angle difference $\phi_1-\phi_2$ is far from an integer multiple of $\pi/2$. The dependence of the errors on $\phi_1-\phi_2$ is described in the text. 
\label{comparison} }
\end{center}
\end{figure}

Figures \ref{quarter_pi} show spatiotemporal distributions of $S_x$ and $S_z$ with the angle difference $\phi_1-\phi_2=\pi/3$. This result shows that the domain pair moves at a constant velocity with the $x$ component of the spin $S_x$ rotating and that the velocity and frequency with the narrower domain are larger than that with the wider one. This is consistent with the analytical result. To compare the numerical results with the analytical one quantitatively, we compute following quantities:
\begin{eqnarray}
{\rm Error}_{\rm vel} = \frac{V_{\rm Num}-V(l)}{V(l)} \times 100,   \label{NU1}
\end{eqnarray}
\begin{eqnarray}
{\rm Error}_{\rm fre} = \frac{\Omega_{\rm Num}-\Omega(l)}{\Omega(l)}\times 100.  \label{NU2}
\end{eqnarray}
Figure \ref{comparison} shows results for these quantities, which demonstrate that ${\rm Error}_{\rm vel} $ and ${\rm Error}_{\rm fre}$ are small as long as the conditions (i) and (ii) are satisfied. 

Note that, if $\phi_1-\phi_2$ is small, ${\rm Error}_{\rm vel}$ is large but ${\rm Error}_{\rm fre}$ is small. 
This is due to Eqs.~(\ref{LL18}) and (\ref{LL18_1}). In this case, $H(X_2,\phi_1-\phi_2)$ in Eq.~(\ref{LL18}) is much smaller than unity but $G(X_2,\phi_1-\phi_2)$ in Eq.~(\ref{LL18_1}) is the order of unity. Then, the leading term for the imaginary part of $\epsilon (x,t)$ in Eq.~(\ref{LL21}), which is related to $V (l)$, is smaller than the order of $e^{-l}$. As shown in the derivation, Eqs.~(\ref{LL32}) and (\ref{LL32_1}) are correct to the order of $e^{-l}$, so that ${\rm Error}_{\rm vel}$ becomes large when the angle difference $\phi_1-\phi_2$ is much smaller than unity. 
In contrast to this case, ${\rm Error}_{\rm vel}$ is small but ${\rm Error}_{\rm fre}$ is large when $\phi_1-\phi_2$ is near $\pi/2$. This is because $H(X_2,\phi_1-\phi_2)$ in Eq.~(\ref{LL18}) is the order of unity but $G(X_2,\phi_1-\phi_2)$ in Eq.~(\ref{LL18_1}) is much smaller than unity.

\thispagestyle{myheadings}


\begin{thebibliography}{99}
\bibitem{comment1} Strictly speaking, the interaction between the domain walls plays a role even in open dissipative systems. However, the effect of dissipation is crucial, and the domain-wall dynamics is driven by the interplay between the interaction and dissipation. Thus, the dynamics is essentially different from the one induced only by the interaction in isolated systems.
\bibitem{PP} We use two properties of the Poisson process: (1) collision events are rare, and (2) collisions occur independently. Furthermore, under these assumptions, the domain-wall pairs are considered to move as separate free particles, which implies that the pairs are well separated.
\bibitem{reason} This is because a 2D domain wall has a curvature $1/R$ that leads to a driving force proportional to $1/R^n$ with the power exponent $n$ and dissipation due to the environment causes self-contraction.
\bibitem{SpRe} Figure \ref{domain_size_exp} shows that the correlation length just after the quench ($t/\tau \sim 800$) is $13.8a_{\rm exp} \sim 5.34 {\rm \mu m}$. By numerically calculating Eq.~({\ref{domain_law3}}) until $9~{\rm s} ~(t/\tau \sim 57000)$, the correlation length increases up to $23.3a_{\rm exp} \sim 9.02 ~{\rm \mu m}$. The difference is about 3.68 ${\rm \mu m}$, so that we can obtain at least 4 data points if the spatial resolution is $1~{\rm \mu m}$.
\bibitem{TBL} The three-body parameter of $^{7} \rm Li$ is not reported in literature. For $^{87}{\rm Rb}$ and  $^{23}{\rm Na}$, the parameters are $4.3 \times 10^{-29} {\rm cm^6/s}$ and $6.6 \times 10^{-30} {\rm cm^6/s}$ for a noncondensed cloud \cite{TBL1}. In the BEC cases, the values are 6 times lower than the non-condensed cases. Generally, an atom with a small radius has a small three-body parameter, so that we speculate that the parameter of $^{7} \rm Li$ is smaller than $10^{-30} {\rm cm^6/s}$.
\end{thebibliography}
\end{document}